\documentclass[conference]{IEEEtran}
\IEEEoverridecommandlockouts
\usepackage[switch]{lineno}

\usepackage[table]{xcolor}
\usepackage{arydshln}
\usepackage{url,tikz}
\usepackage{algorithm,algpseudocode,amsmath,amssymb,amsfonts}
\usepackage{balance,booktabs,cite,fixme,graphicx,hyperref}
\usepackage{caption}
\usepackage{textcomp,xcolor}

\usepackage{ipdps26repro}
\fxsetup{
  status=draft,
  theme=color
}
\def\BibTeX{{\rm B\kern-.05em{\sc i\kern-.025em b}\kern-.08em
    T\kern-.1667em\lower.7ex\hbox{E}\kern-.125emX}}

\begin{document}

%%%Comments
% \newcommand{\Emmanuel}[1]{\textcolor{red}{{\bf [Emmanuel: #1]}}}
% \newcommand{\ritvik}[1]{\textcolor{blue}{{\bf [Ritvik: #1]}}}

\newif\ifreview
\reviewfalse
\newcommand{\review}[1]{\ifreview\textcolor{blue}{#1}\else#1\fi}

\definecolor{babcolor}{rgb}{0.9,0.45,0.1}
\definecolor{myyellow}{RGB}{255,204,0}
\newcommand{\wahib}[1] {{\color{babcolor}{#1}}}
\DeclareRobustCommand*\circled[1]{\tikz[baseline=(char.base)]{
            \node[shape=circle,fill,inner sep=0.5pt] (char) {\textcolor{white}{#1}};}}

\DeclareRobustCommand*\circledred[1]{\tikz[baseline=(char.base)]{
  \node[shape=circle, fill=red!80!black, inner sep=0.5pt] (char) {\textcolor{white}{#1}};
}}

\DeclareRobustCommand*\circledblack[1]{\tikz[baseline=(char.base)]{
  \node[shape=circle, fill=black!80!black, inner sep=0.5pt] (char) {\textcolor{white}{#1}};
}}

\DeclareRobustCommand*\squaredgreen[1]{\tikz[baseline=(char.base)]{
  \node[shape=rectangle, fill=green!80!black, inner sep=0.8pt] (char) {\textcolor{white}{#1}};
}}

\DeclareRobustCommand*\squaredblack[1]{\tikz[baseline=(char.base)]{
  \node[shape=rectangle, fill=black, inner sep=0.8pt] (char) {\textcolor{white}{#1}};
}}

\DeclareRobustCommand*\ocis[1]{\tikz[baseline=(char.base)]{
  \node[shape=rectangle, draw=orange!90!black, inner sep=0.8pt] (char)
    {\tikz[baseline=(n.base)]\node[shape=circle, fill=orange!90!black, inner sep=0.5pt] (n) {\textcolor{white}{#1}};};
}}

\DeclareRobustCommand*\blueOcis[1]{\tikz[baseline=(char.base)]{
  \node[shape=rectangle, draw=blue!90!black, inner sep=0.8pt] (char)
    {\tikz[baseline=(n.base)]\node[shape=circle, fill=blue!90!black, inner sep=0.5pt] (n) {\textcolor{white}{#1}};};
}}

%\title{Cancer Detection using Pruned Combinatorial Combinations of Genetic Mutations
%}
\title{\review{Looking for (Genomic) Needles in a Haystack: \\ Sparsity-Driven Search for Identifying Correlated Genetic Mutations in Cancer}}
%\title{Sparsity-Driven Search for Identifying Correlated Genetic Mutations in Cancer}

% \author{
%   \IEEEauthorblockN{
%     Ritvik Prabhu\IEEEauthorrefmark{1},
%     Emil Vatai\IEEEauthorrefmark{2}\IEEEauthorrefmark{4}\thanks{\IEEEauthorrefmark{4} Emil Vatai and Bernard Moussad contributed equally to this work.},
%     Bernard Moussad\IEEEauthorrefmark{1}\IEEEauthorrefmark{4},
%     Emmanuel Jeannot\IEEEauthorrefmark{2},
%     Wu-chun Feng\IEEEauthorrefmark{1} and
%     Mohamed Wahib\IEEEauthorrefmark{2}
%   }
%   \IEEEauthorblockA{\IEEEauthorrefmark{1}\textit{Department of Computer Science} \\
%     \textit{Virginia Tech}\\
%     Blacksburg, United States of America \\
%     Email: \{ritvikp,bernardm,wfeng\}@vt.edu}
%   \IEEEauthorblockA{\IEEEauthorrefmark{2}\textit{High Performance Artificial Intelligence Systems Research Team} \\
%     \textit{RIKEN Center for Computational Science}\\
%     Kobe, Japan \\
%     Email: \{emil.vatai,mohamed.attia\}@riken.jp, ejeannot@ddn.com}
% }

\author{
  \IEEEauthorblockN{
    Ritvik Prabhu\IEEEauthorrefmark{1},
    Emil Vatai\IEEEauthorrefmark{2}\IEEEauthorrefmark{5}\thanks{\IEEEauthorrefmark{5} Emil Vatai and Bernard Moussad contributed equally to this work.},
    Bernard Moussad\IEEEauthorrefmark{1}\IEEEauthorrefmark{5},
    Emmanuel Jeannot\IEEEauthorrefmark{4},\\
    Ramu Anandakrishnan\IEEEauthorrefmark{3},
    Wu-chun Feng\IEEEauthorrefmark{1}, 
    Mohamed Wahib\IEEEauthorrefmark{2}
  }
  \vspace*{4pt} 
  \IEEEauthorblockA{\IEEEauthorrefmark{1}\textit{Department of Computer Science}, \textit{Virginia Tech}, Blacksburg, USA\\
  \IEEEauthorrefmark{3}\textit{Biomedical Sciences}, \textit{Edward Via College of Osteopathic Medicine}, Blacksburg, USA\\
  Email: \{ritvikp,bernardm,ramu,wfeng\}@vt.edu}
  %\IEEEauthorblockA{\IEEEauthorrefmark{3}\textit{Biomedical Sciences}, \textit{Edward Via College of Osteopathic Medicine}, Blacksburg, USA\\
  %Email: ramu@vt.edu}
  
\IEEEauthorblockA{\IEEEauthorrefmark{4} \textit{Inria, Univ. Bordeaux, LaBRI}, Talence, France\\
  Email: emmanuel.jeannot@inria.fr}
  
  \IEEEauthorblockA{\IEEEauthorrefmark{2} \textit{RIKEN Center for Computational Science}, Kobe, Japan\\
  Email: \{emil.vatai,mohamed.attia\}@riken.jp}

}

\ifCLASSOPTIONpeerreview
  \linenumbers
\IEEEpeerreviewmaketitle
\else
\maketitle %%IEEE%%
\fi

\begin{abstract}
Cancer typically arises not from a single genetic mutation (i.e., hit) but from multi-hit combinations that accumulate within cells. 
%Identifying these higher-order hits has the potential to transform our understanding of tumor development and guide precisely targeted combination therapies. 
However, enumerating multi-hit 
%sets 
combinations becomes exponentially more expensive computationally 
%explosive 
as the number of candidate hit gene combinations grow, i.e. %$\binom{20{,}000}{h}$
$C^{20,000}_{h}$, where $20,000$ is the number of genes in the human genome and $h$ is the number of hits.
To address this challenge, we present an algorithmic framework, called Pruned Depth-First Search (P-DFS) that leverages the high \emph{sparsity} in tumor mutation data to prune large portions of the search space. 
% Specifically, P-DFS, which is grounded in a \emph{depth-first search backtracking} technique, prunes infeasible gene subsets early, while a \review{standard greedy} \emph{weighted set cover} (WSC) formulation systematically scores and selects the most discriminative combinations. 
Specifically, P-DFS \review{(the main contribution of this paper) - a pruning technique that exploits sparsity to drastically reduce the otherwise exponential $h$-hit search space for candidate combinations used by Weighted Set Cover -} which is grounded in a \emph{depth-first search backtracking} technique, prunes infeasible gene subsets early, while a \review{standard greedy} \emph{weighted set cover} (WSC) formulation systematically scores and selects the most discriminative combinations.

By intertwining these ideas with optimized bitwise operations and a scalable
%master-worker paradigm 
distributed algorithm on high-performance computing clusters, 
our algorithm can achieve $\approx$ 90 - 98\% reduction in visited combinations for 4-hits, and roughly a $183~\times$ speedup over the exhaustive set cover approach (which is algorithmically NP-complete) measured on 147{,}456 ranks.
% we massively improve the execution time, 
% %reduce runtimes 
% relative to the exhaustive set cover approach, which is algorithmically NP-complete.
%naive enumeration methods. 
In doing so, our method can feasibly handle four-hit and even higher-order gene hits, achieving both speed and resource efficiency.
%qualitative improvements (discovering mutation combinations previously deemed intractable) and quantitative gains in speed and resource efficiency.
\end{abstract}
% \begin{IEEEkeywords}
% Cancer genomics, somatic mutations, multi-hit mutation combinations, weighted set cover, combinatorial optimization, sparsity-aware pruning, depth-first search, MPI, work stealing, load balancing
% \end{IEEEkeywords}
%\begin{IEEEkeywords}
%component, formatting, style, styling, insert
%\end{IEEEkeywords}

\section{Introduction}

\begin{figure*}[t]
  \centering
  \includegraphics[width=1\textwidth]{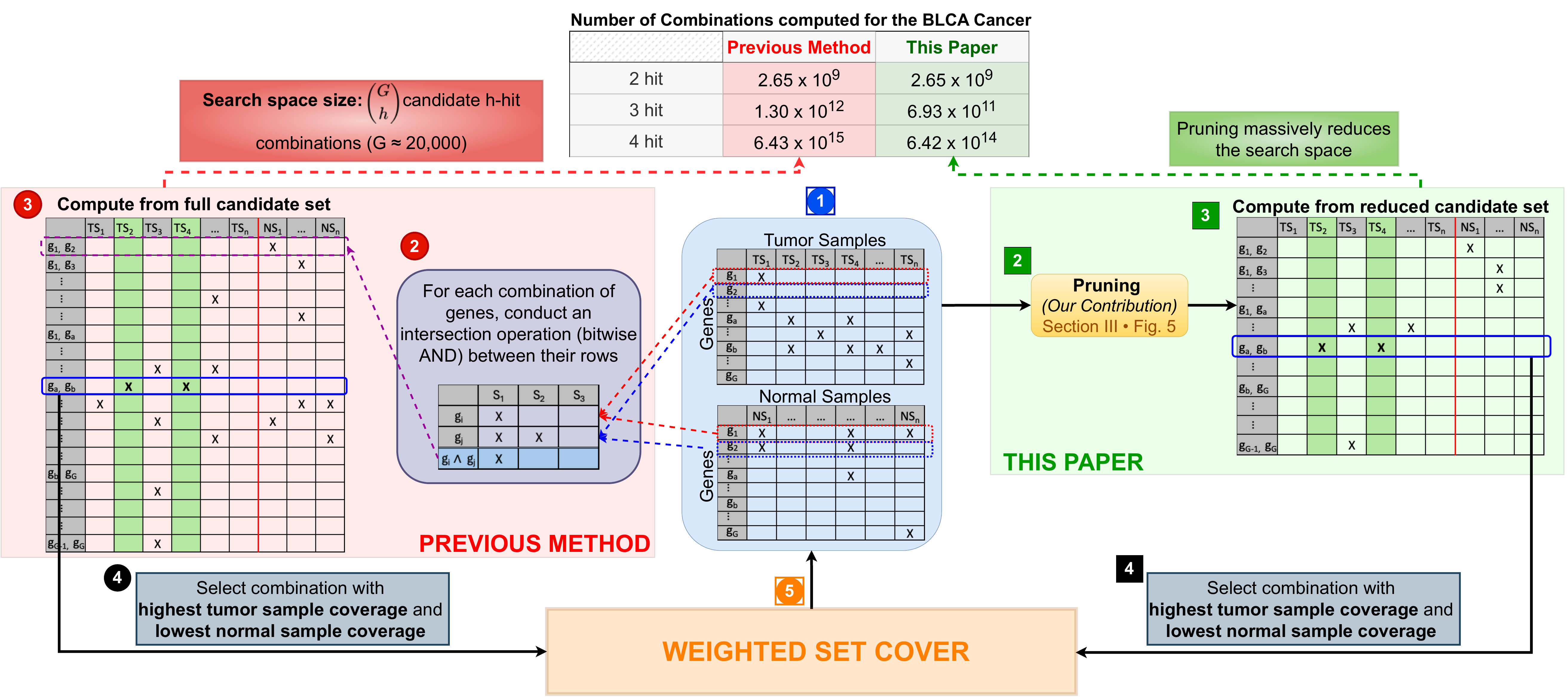}
\caption{\review{
\textbf{Overview of workflow.}
\blueOcis{1} \textbf{Input:} tumor and normal mutation matrices, represented as per-gene bitsets over samples.
\circledred{2}--\circledred{3} \textbf{Previous method (exhaustive enumeration):} for \emph{every} $h$-hit gene combination, the prior pipeline performs a bitwise AND across the corresponding gene rows (\circledred{2}) and computes over the full candidate table (\circledred{3}), which quickly becomes intractable as $h$ grows.
\squaredgreen{2}--\squaredgreen{3} \textbf{This paper (prune-then-evaluate):} our main contribution is P-DFS, a sparsity-driven \underline{p}runed \underline{d}epth-\underline{f}irst \underline{s}earch
%pruning stage 
that \emph{massively prunes} the $h$-hit search space (\squaredgreen{2}) before scoring (see \S\ref{sec:algorithm} and Fig.~\ref{fig:new-workflow}).
This produces a substantially smaller candidate table to evaluate (\squaredgreen{3}), yielding significant compute-time savings by avoiding bitwise-AND intersections for the vast majority of combinations.
\circledblack{4} and \squaredblack{4} \textbf{Selection step (common to both):} choose the candidate combination with highest tumor coverage and lowest normal coverage, and pass it into the weighted set cover procedure \ocis{5}.
The table at the top summarizes the resulting reduction in combinations explored, which becomes more pronounced as the number of hits $h$ increases.
}}
\label{fig:high-level-wsc}
\end{figure*}

Genetic mutations occur routinely in our cells, yet most of these mutations are benign. Occasionally, however, a specific subset of mutations arises that can pave the way for cancer development. Cancer itself is not a single disease, rather, it is a diverse group of disorders defined by unchecked cell growth~\cite{brown2023updating}. Despite decades of effort, it remains the second leading cause of death worldwide, projected to claim around 618,120 lives in the United States alone in 2025~\cite{siegel2025cancer}. The crux of the challenge lies in the fact that cancer typically does not emerge from a single genetic mutation (“hit”) but rather from precise, multi-hit combinations of a small number (typically between 2--9 hits) of genetic alterations~\cite{anandakrishnan2019estimating}.

Understanding these multi-hit patterns has profound implications. First, it illustrates how cancer starts and progresses, providing insight into the biological mechanisms behind tumor formation. Second, identifying the exact sets of mutations responsible for a given case of cancer can guide the design of more targeted combination therapies~\cite{al-lazikani_combinatorial_2012, ledford_cocktails_2016}. Nevertheless, pinpointing these critical mutation sets is akin to searching for needles in a genomic haystack. With roughly 20,000 genes in the human genome~\cite{international_human_genome_sequencing_consortium_finishing_2004}, even two-hit combinations yield on the order of $2 \times 10^8$ possibilities, while four-hit and five-hit combinations balloon into the range of $10^{15}$ to $10^{19}$, respectively. Exhaustively exploring all such combinations quickly overwhelms traditional computational approaches; %e.g., 
%For example, 
recent estimates suggest that identifying all four-hit combinations would require over 500 years on a single CPU~\cite{dash_scaling_2021}.

To tackle this challenge, multi-hit combination analysis has emerged as a prominent strategy.
At its core, this approach aims to pinpoint which specific combinations of genetic mutations are most relevant to carcinogenesis, i.e.\ the process by which normal cells transform into cancer cells.
One leading paradigm views the task as a weighted set cover (WSC) problem, an NP-complete problem, where each candidate mutation set ``covers'' the tumor samples carrying that set and ideally excludes normal samples~\cite{al2020identifying}.
Although WSC can be highly accurate, all possible gene combinations must be exhaustively enumerated, an endeavor that grows exponentially with each additional ``hit.''
Even powerful high-performance computing (HPC) systems struggle to search beyond three-hit or four-hit mutation sets without encountering prohibitively large runtimes or system limitations like I/O bottlenecks, memory constraints, and communication overhead~\cite{dash_scaling_2021}.

To cope with the exponential explosion of the search space, researchers have explored heuristics that avoid enumerating the entire combination space. One such approach, BiGPICC, quickly identifies higher-order hits by leveraging a graph-based method. However, BiGPICC omits large swaths of the combination space in the name of speed, which can degrade classification accuracy~\cite{oles_bigpicc_2023}. In practice, then, the cancer-genomics community faces a trade-off: comprehensive methods that handle only a few hits versus heuristic-based approaches that find more hits but risk missing the precise “needles” in the genomic haystack.

% In this paper, we introduce a novel HPC-driven framework to push the boundary of multi-hit analyses. We build upon the rigor of WSC-based methodology but overcome its typical limitations through the utilization of advanced supercomputing resources and optimizations that capitalize on the sparsity inherent to carcinogenesis data. By doing so, our approach navigates the massive search space that was previously considered intractable. Specifically, we introduce a new algorithm, \emph{pruned depth-first search} (P-DFS) for traversing the search space of multi-hit combinations. In particular, we leverage the high sparsity in gene mutations to do a distributed depth-first search with backtracking to prune the combinatorial search space, thus dramatically improves the efficiency.

In this paper, we introduce a novel HPC-driven framework to push the boundary of multi-hit analyses. We build upon the rigor of WSC-based methodology but overcome its typical limitations through the utilization of advanced supercomputing resources and optimizations that capitalize on the sparsity inherent to carcinogenesis data. By doing so, our approach navigates the massive search space that was previously considered intractable. Specifically, we introduce a new algorithm, \emph{pruned depth-first search} (P-DFS) for traversing the search space of multi-hit combinations\review{---our main contribution, which reduces the otherwise exhaustive candidate search space that dominates WSC-based pipelines (Fig.~\ref{fig:high-level-wsc})}. In particular, we leverage the high sparsity in gene mutations to do a distributed depth-first search with backtracking to prune the combinatorial search space, \review{thereby shrinking the candidate set before applying the \emph{standard greedy} WSC selection step,} thus dramatically \review{improving} the efficiency.

% our contributions are:
%that leverages Fugaku, the world's fastest homogeneous supercomputer,
% \begin{itemize}
%     \item \wahib{We introduce a new algorithm for searching the search the space of multi-hit combinations. In particular, we leverage the high sparsity in gene mutations to do a distributed depth-first search with backtracking to prune the combinatorial search space, which dramatically improves the efficiency.} 
% \end{itemize}

Through this effort, we seek not only to advance our understanding of the genetics underlying cancer but also to provide a practical path toward precisely tailored combination therapies that could improve outcomes for patients worldwide.

The remainder of the paper proceeds as follows. In %Section~\ref{sec:background}, 
\S\ref{sec:background}, we review background and related work on multi-hit modeling and search formulations. 
%Section~\ref{sec:algorithm} 
\S\ref{sec:algorithm} presents our pruned depth-first search (P-DFS) method, including sparsity-aware preprocessing, the core pruning algorithm, and topology-aware work distribution with hierarchical collectives. %Section~\ref{sec:results} 
\S\ref{sec:results} reports the results on the Fugaku supercomputer, %which include 
including datasets, pruning efficiency, accuracy on held-out data, load balancing with work stealing, and strong scaling. %Section~\ref{sec:conclusion} 
\S\ref{sec:conclusion} provides a summary of our findings.

\section{Background and Related Work}
\label{sec:background}

\subsection{Historical Foundations of the Multi-Hit Concept}

A seminal moment in understanding cancer's genetic basis came from Alfred Knudson's 1971 ``two-hit hypothesis,'' which suggested that certain cancers, specifically retinoblastoma, emerge only after mutations have occurred in both copies of a tumor suppressor gene~\cite{knudson_mutation_1971}. While Knudson focused on a rare pediatric tumor, his ideas gave rise to a wave of research showing that various solid tumors (e.g., breast, colon, and colorectal cancers) likely require multiple genetic ``hits'' to shift from normal cellular regulation to malignant growth~\cite{zhang_estimating_2005, tomasetti_only_2015, little_stochastic_2003}.

Over the following decades, scientists increasingly recognized that no single ``magic number'' of hits applies universally across all tumor types. This led to a deeper statistical and mathematical modeling of carcinogenesis. Early modeling efforts often assumed relatively simple, uniform mutation rates across genes~\cite{bavarva_dynamic_2014, wu_robust_2009, alexandrov_clock-like_2015, paashuis-lew_spontaneous_1998}, but they could not account for the variability in how fast or slow different genes accumulate alterations. Building upon these limitations, Anandakrishnan et al. introduced a more data-driven model that focuses on the observed distribution of somatic mutations rather than on any single pre-specified rate~\cite{anandakrishnan2019estimating}. They established that many cancers seem to arise from combinations of between 2 and 9 hits, depending on tumor type.

\begin{figure}[t]
    \centering
    \includegraphics[width=0.5\textwidth,height=0.5\textheight,keepaspectratio]{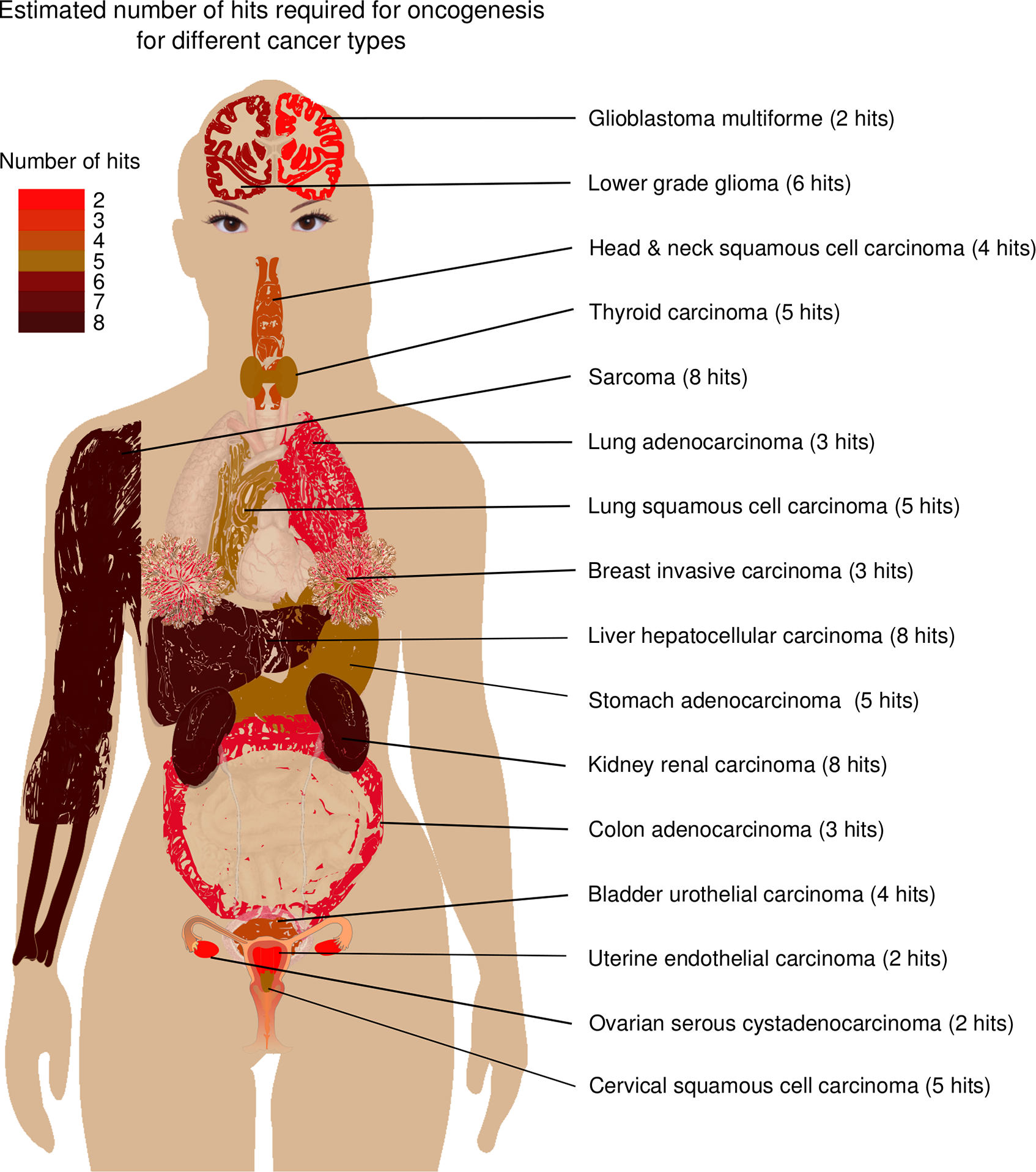}
    \caption{Visualization of estimated hit numbers for each cancer type from \cite{anandakrishnan2019estimating}. Derived from the public domain image by M. Haggstrom (2014)~\cite{wiki:Hag}.}
    \label{fig:hit-numbers}
\end{figure}

%\noindent 
As shown in Fig.~\ref{fig:hit-numbers}, the number of hits required can vary, reinforcing the idea that cancer is not a single entity but rather a multitude of genetic scenarios each leading to a common endpoint of tumor formation. This insight highlights the importance of developing comprehensive computational tools capable of sifting through a large and heterogeneous genomic search space.

\subsection{Approaches for Identifying Multi-Hit Combinations}

From a computer science perspective, the multi-hit problem can be formulated as follows. Given a binary mutation matrix $X \in \{ 0, 1 \}^{G \times n}$, where $G$ is the total number of genes under consideration and $n$ is the number of samples (from patients with and without cancer, to which we refer to as tumor samples and normal samples, respectively).
Each row corresponds to a gene and each column to a sample, i.e., $x_{g,s}$, the element in the $g$-th row and $s$-th column of $X$ is set to 1 if the sample $s$ has a mutation in the gene $g$ (and 0 otherwise).
We say a combination $\gamma=\{ g_1,\ldots,g_h \}$ of $h$ genes, ``covers'' a particular sample $s$ if \emph{all} genes are mutated in sample~$s$, i.e., $X_{g_1,s} = X_{g_2,s} = \cdots = X_{g_h,s} = 1$.
The set of samples covered by a single combination $\gamma$ is denoted by:
\begin{equation}\label{eq:cover-gamma}
  \mathcal{C}(\gamma) = \{ s : \forall g \in \gamma, x_{g, s} = 1 \},
\end{equation}
while the set of samples covered by a set of combinations $\Gamma$ is the union of the samples covered by the combinations in the set \(\mathcal{C}(\Gamma) = \cup_{\gamma \in \Gamma} \mathcal{C}(\gamma).
\)
% \fxnote*{This sentence is redundant. -- Emil}{Thus, a candidate set of $h$ genes (or "hits") covers a sample if and only if all $h$ genes are mutated in that sample.}
% \fxnote*{I don't get this sentence -- Emil}{Each such set therefore corresponds to the set of samples in which these mutations co-occur.}

As large-scale genomic data became widely accessible (spanning tens of thousands of genes in thousands of samples), finding the specific gene sets that constitute these multi-hit combinations became a computationally daunting task. Two primary solutions have emerged to address this challenge: (i) methods grounded in the weighted set cover (WSC) problem and (ii) graph-based approaches.

\subsubsection{Weighted Set Cover (WSC) Paradigm}

One of the most prominent solution strategies for multi-hit analysis is to formulate the problem as a weighted set cover (WSC)~\cite{dash_differentiating_2019, al2020identifying, dash_scaling_2021}.
Here, the aim is to find a set of combinations with minimal weights that cover all tumor samples.  % It is MIN. The greedy step is MAX
That is, each combination $\gamma$ has a weight $F(\gamma)$ associated with it, and the goal is to find the set of combinations $\Gamma^*$ that has minimal combined weight among the sets that also covers the tumor samples.

Formally, denoting the set of all \(h\) gene combinations with $\Gamma_h$,  the minimality objective can be stated as
\begin{equation}\label{eq:goal}
  F(\Gamma^{*}) 
  = \underset{\Gamma \subset \Gamma_h}\min\{  F(\Gamma)\}
  \quad \text{i.e.,} \quad 
  \Gamma^{*} 
  = \underset{\Gamma \subset \Gamma_h}{\operatorname{argmin}}\{F(\Gamma)\}
\end{equation}
while \(T \subset \mathcal{C}(\Gamma^*)\) expresses the fact that $\Gamma^*$ covers the tumor samples, $T$ (where, \(\mathcal{C}(\Gamma) = \cup_{\gamma \in \Gamma}\mathcal{C}(\gamma)\)). 
% Here, each combination of $h$ genes is treated as a “set” that \emph{covers} all tumor samples in which those $h$ genes co-occur.
% The coverage is then assigned a weight based on how well the $h$-gene combination distinguishes tumor samples from normal samples~\cite{dash_differentiating_2019, al2020identifying, dash_scaling_2021}.

\paragraph{Scoring and Algorithmic Steps}
% \fxnote{This part is highly modified and some cut parts are left in the comments}
As an example of a WSC-based method, Dash et al.~\cite{dash_differentiating_2019} maps the multi-hit problem to WSC and applies a greedy strategy to find sets of genes that appear \emph{frequently} in tumor samples but \emph{rarely} in normal samples.
Formally, let $N_t(X)$ and $N_n(X)$ be the total number of tumor and normal samples, respectively.
For each $h$-hit gene combination $\gamma$, $T_{+}(\gamma, X)$ is the number of tumor samples that contain \emph{all} $g \in \gamma$ genes (true positives), and $T_{-}(\gamma, X)$ the number of normal samples that do \emph{not} contain any of the $h$ genes in the combination (true negatives).

A commonly used objective or “weight” function is~\cite{dash_differentiating_2019}:
\[
F(\gamma, X) = \frac{\alpha \, T_{+}(\gamma, X) + T_{-}(\gamma, X)}{N_t(X) + N_n(X)},
\]
where $\alpha$ (often set to $0.1$) penalizes the approach's inherent bias toward maximizing $T_{+}$ over $T_{-}$.
Until all tumor samples are covered, the algorithm, greedily selects the $h$-hit combination $\gamma^{*}$ with the highest $F(\gamma, X)$ value, and removes the samples (i.e., columns) from $X$ which are covered by $\gamma^{*}$ as described by Algorithm~\ref{alg:greedy}.
\begin{algorithm}[t]
\caption{Greedy Set Cover}\label{alg:greedy}
\begin{algorithmic}[1] % [1] adds line numbering
\Require $X$ gene mutation matrix, $T$ indices of tumor samples, $h$ number of hits
\Ensure $\Gamma$ covers $T$, i.e., $T \subset \mathcal{C}(\Gamma)$
\State $C \gets \emptyset$ \Comment{Set of covered samples}
\State $\Gamma \gets \emptyset$ \Comment{The variable holding the results}
\While{$C \subsetneq T$}\Comment{Are all tumors samples covered?}
    \State $\gamma^{*} \gets \underset{\gamma \in \Gamma_h}{\operatorname{argmax}} F(\gamma, X)$ \Comment{Get the best combination}\label{stp:f}
    \State $\Gamma \gets \Gamma \cup \{ \gamma^{*} \}$ \Comment{Append $\gamma^{*}$ to the results}
    \State $C \gets C \cup \mathcal{C}(\gamma^{*})$ \Comment{Update the set of covered samples}
    \State $X \gets \textsc{DelCols}(X, \gamma^{*})$ \Comment{Remove samples from $X$}
\EndWhile
\State\Return $\Gamma$
\end{algorithmic}
\end{algorithm}

% The algorithm then proceeds in an iterative fashion: \ritvik{Should we make a visual for this? Maybe show some code?}
% \begin{enumerate}
%     \item Enumerate all $h$-gene combinations and compute $F$ for each.
%     \item Select the combination that achieves the highest $F$.
%     \item Exclude (i.e., “cover”) all tumor samples hit by that gene combination and repeat until no tumor samples remain uncovered.
% \end{enumerate}

% Hence, each iteration filters out certain tumors from further consideration and focuses subsequent searches on increasingly smaller subsets of tumors that remain uncovered.

\paragraph{Computational Complexity}\label{sec:complexity}
While straightforward in concept, enumerating \emph{all} possible $h$-gene sets in line~\ref{stp:f} of Algorithm~\ref{alg:greedy} is computationally costly.
The total number of $h$-gene combinations is $M = \binom{G}{h}$, where $G$ is on the order of 20,000 genes. Even for $h=4$, $M \approx 7 \times 10^{15}$, making an exhaustive search prohibitive. Formally, WSC is an NP-complete problem with no known polynomial-time exact solution~\cite{dash_scaling_2021}. Approximate or greedy algorithms reduce the worst-case complexity to $O(M \times N_t)$, which is still tractable only for small $h$. Thus, early multi-hit approaches using WSC focused on 2-hit or 3-hit gene sets, and more recent GPU-accelerated methods extended these analyses up to 5 hits~\cite{al2020identifying, dash_scaling_2021, dash_distributing_2023}. However, jumping to 6-hit combinations (much less 8 or 9) remains largely infeasible via exhaustive enumeration or naive greedy methods.

\paragraph{Pruned and Parallel DFS for Candidate Generation}
A common way to mitigate the $\binom{G}{h}$ blowup is to treat $\Gamma_h$ as an implicit search tree and traverse it with depth-first search (DFS), while pruning subtrees that cannot yield useful candidates.
This idea is well established in frequent itemset mining: DFS-style enumeration of $h$-way co-occurrences is paired with anti-monotone bounds to prune away large portions of the candidate lattice~\cite{borgelt_2012_frequent_item_set_mining, uno_etal_2003_lcm, luna_etal_2019_fim_review}.
At scale, these DFS/backtracking trees are routinely parallelized by partitioning the search space (often by prefixes) and using dynamic load balancing like randomized work donation or work stealing, to handle highly irregular subtree sizes~\cite{karp_zhang_1993_backtrack, blumofe_leiserson_1999_workstealing, hiraishi_etal_2009_backtracking_load_balancing}.
In the same spirit, our approach uses a pruned, parallel DFS to generate and score only promising $h$-hit candidates, rather than explicitly materializing all of $\Gamma_h$. Although exact set-cover solvers can use Branch and Bound (B\&B) with relaxation-based bounds for safe pruning, we do not use B\&B because our WSC-style greedy routine does not maintain such bounds~\cite{caprara_etal_2000_set_covering}.

\subsubsection{Graph-Based Representations}

While WSC aims to enumerate combinations explicitly, graph-based methods, such as BiGPICC~\cite{oles_bigpicc_2023}, represent genomic data in the form of bipartite networks with gene nodes on one side and sample nodes on the other. Edges exist only if a gene is mutated in a given sample. Clustering algorithms (like the Leiden method~\cite{traag_louvain_2019}) then aim to group genes that frequently co-occur across patients.

\paragraph{Speed vs.\ Coverage}
Graph-based strategies often reduce computational costs compared to full or partial enumerations. By clustering co-mutated genes, they can efficiently highlight general mutation patterns across large datasets. However, the trade-off is that crucial mutations may be missed when aggregated into clusters, which as a result, affects the classification accuracy as compared to the WSC approach. Thus, graph-based solutions are often preferred for quick large-scale screening or when fewer computational resources are available, but they may miss subtle ``needle-in-a-haystack'' multi-hit combinations.

% \subsection{Pruning in Combinatorial Tree Search}
% Most of the runtime in our workflow is spent exploring the combinatorial tree of $h$-hit gene combinations, so performance depends on pruning branches as early as possible. In tree search, once a partial assignment is infeasible, the subtree below it can be discarded. Backtracking with pruning rules is a standard way to avoid enumerating all candidate solutions \cite{BitnerReingold1975}.
% Infeasibility is not the only reason to prune. Dominance relations can also justify skipping a subtree when one partial state is provably no better than another for all possible completions \cite{Ibaraki1977DominanceBB}.
% Branch-and-bound adds another pruning rule that uses objective bounds. A node can be discarded if its bound cannot beat the current best solution \cite{Smith1984RandomTreesBB,FukunagaNarendra1975kNNBB}.
% Our pruning, however, uses only sparsity-based feasibility (empty intersections under bitset operations) and never computes objective bounds for the WSC algorithm, so our problem cannot be viewed as a branch-and-bound method for WSC.

\section{Pruned Depth-First Search (P-DFS): Pruning Algorithm for $h$-hit Gene Mutations}\label{sec:algorithm}

% \subsection{Overview}
% \label{sec:algo-overview}

% \subsection{Motivation}
% \label{sec:algo-motivation}

% Our central objective is to discover multi-hit gene combinations without compromising accuracy. To that end, we prioritize the WSC algorithm over purely graph-based alternatives.

% A second motivation is the \emph{extreme sparsity} of cancer genomic data. In typical somatic-mutation matrices, most entries are zero: each tumor sample has mutations in only a small subset of genes, and each gene is altered in only a small subset of tumors. In our datasets, the mutation matrix has median sparsity \textbf{95.61\%} as seen in Fig.~\ref{fig:sparsity-violin}. This motivates methods that exploits this sparsity.

% \begin{figure}[t]
%     \centering
%     \includegraphics[width=\linewidth]{figs/violin_sparsity_all.png}
%     \caption{Sparsity of the carcinogenesis mutation matrix datasets used in this study. Median: \textbf{95.61\%}. High sparsity underscores the rarity of mutations in the datasets.}
%     \label{fig:sparsity-violin}
% \end{figure}

% Our motivation, therefore, is to preserve WSC's accuracy benefits while engineering the computation to be more computationally tractable at scale \fxnote*{Added by Emil}{while exploiting sparsity}.

\subsection{Motivation}
\label{sec:algo-motivation}

% Our goal is to discover multi-hit gene combinations without sacrificing accuracy. We therefore build on the WSC paradigm, which directly optimizes tumor/normal discrimination rather than relying solely on graph clustering, which may blur the sample-level signal.

Our goal is to discover multi-hit gene combinations efficiently\review{---in a reasonable amount of time---}without sacrificing accuracy. We therefore build on the WSC paradigm, which directly optimizes tumor/normal discrimination rather than relying solely on graph clustering, which may blur the sample-level signal, \review{and focus our contribution on shrinking the otherwise exhaustive $h$-hit candidate search space (via sparsity-driven pruning) while keeping the WSC scoring objective unchanged}.

\begin{figure}[t]
    \centering    \includegraphics[width=\linewidth]{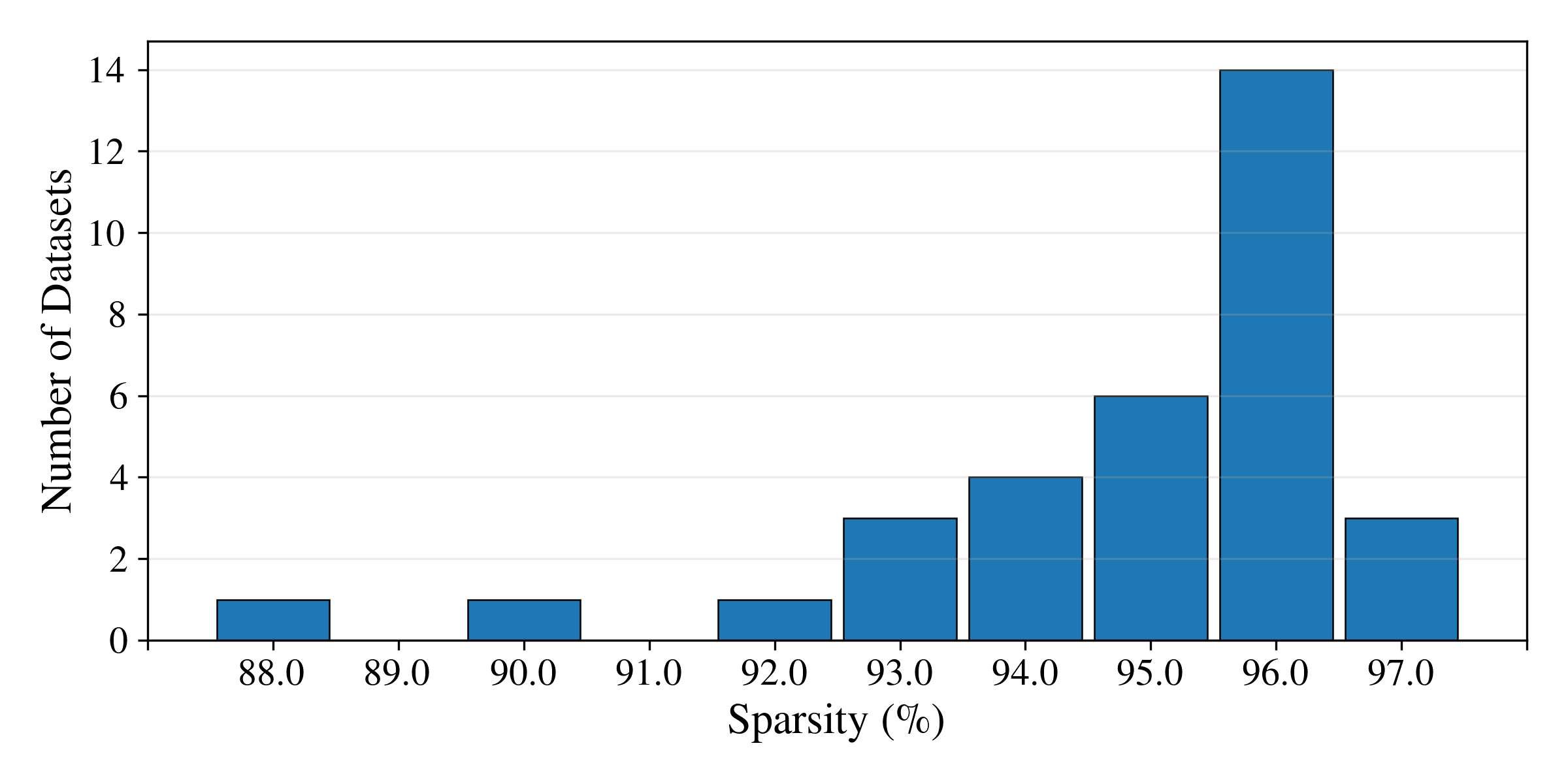}
    \caption{Measured sparsity of all the carcinogenesis mutation matrix datasets available through TCGA. Median: \textbf{95.61\%}. High sparsity underscores the rarity of mutations in all of the cancer datasets.}
    \vspace{15pt}
\label{fig:sparsity-violin}
\end{figure}

The first empirical observation motivating our design is the fact that the main loop in Algorithm~\ref{alg:greedy} is executed at most a tens of times.
More concretely, in datasets studied by prior work, the greedy cover typically terminated in $\le 20$ iterations~\cite{al2020identifying, dash_differentiating_2019}.
The program spends most of its time executing the greedy maximum search in step~\ref{stp:f}, so optimizing this step automatically improves the performance of the whole application.
Another key observation behind the motivation is the extreme sparsity of somatic-mutation matrices. Most entries are zero: each tumor harbors alterations in only a small subset of genes, and each gene is altered in only a small subset of tumors. This “few mountains, many hills’’ profile and pathway-level mutual exclusivity have been extensively documented in cancer genomics \cite{vogelstein2013cancer}. Our datasets follow the same pattern, where the median sparsity is 95.61\% (Fig.~\ref{fig:sparsity-violin}), which immediately suggests algorithmic opportunities.

The challenge in the original WSC approach is the combinatorial growth: the number of $h$-hit candidates scales as ${G \choose h}$, so even modest increases in $h$ cause an explosion in the search space.
Table~\ref{tab:comb-growth} shows the number of $h$-gene candidates, $\binom{G}{h}$ and the corresponding cancers which require that number of gene hits for diagnosis. Even a one-hit increase yields orders of magnitude more combinations, making naive enumeration intractable (calculated here for $G{=}20{,}000$ genes).
Naively enumerating all combinations to retain WSC’s accuracy becomes computationally intractable.

% \begin{figure}[t]
%     \centering
%     \includegraphics[width=\linewidth]{figs/combinatorialExplosion.png}
%     \caption{Combinatorial explosion of the WSC search space. Bars show the number of $h$-gene candidates, $\binom{G}{h}$, versus hits $h$ (log scale). Even a one-hit increase yields orders of magnitude more combinations, making naive enumeration intractable (illustrated here for $G{=}20{,}000$ genes).\wahib{@Ritvik: this would be more readable as a table, not figure.}}
%     \label{fig:comb-growth}
% \end{figure}

\begin{table}[t]
  \renewcommand{\arraystretch}{1.3}
  \caption{
    Combinatorial explosion of the WSC search space.
  }\label{tab:comb-growth}
  \centering
  %%% \resizebox{\columnwidth}{!}{%
  \begin{tabular}{lll}
    \toprule
    Hits & Combinations  & Cancers\\
    \midrule
    2 & $1.99 \times 10^{08}$ & GBM, OV, UCEC\\
    3 & $1.33 \times 10^{12}$ & KIRC, LUDA, SKCM, COAD, BRCA\\
    4 & $6.66 \times 10^{15}$ & BLCA, HNSC\\
    5 & $2.66 \times 10^{19}$ & THCA, STAD, CESC, LUSC\\
    6 & $8.88 \times 10^{22}$ & LGG\\
    7 & $2.53 \times 10^{26}$ & DLBC, PRAD, ESCA\\
    8 & $6.34 \times 10^{29}$ & SARC, KIRP, LIHC\\
    9 & $1.40 \times 10^{33}$ & UCS, UVM, KICH, MESO, AC, PCPG\\
    \bottomrule
  \end{tabular}%
  %%% }
\end{table}

Our approach is to exploit sparsity to prune the search space and hence tame the combinatorial explosion while retaining WSC’s accuracy. Specifically, we structure the computation around early-termination bounds that prune branches as soon as partial intersections go empty. This would leverage the fact that true co-occurrences are rare, so most computation paths die quickly, leaving only a small collection of promising sets to evaluate thoroughly. 

\subsection{Overview}
\label{sec:overview}
\begin{figure*}[t]
    \centering
    \includegraphics[width=\textwidth]{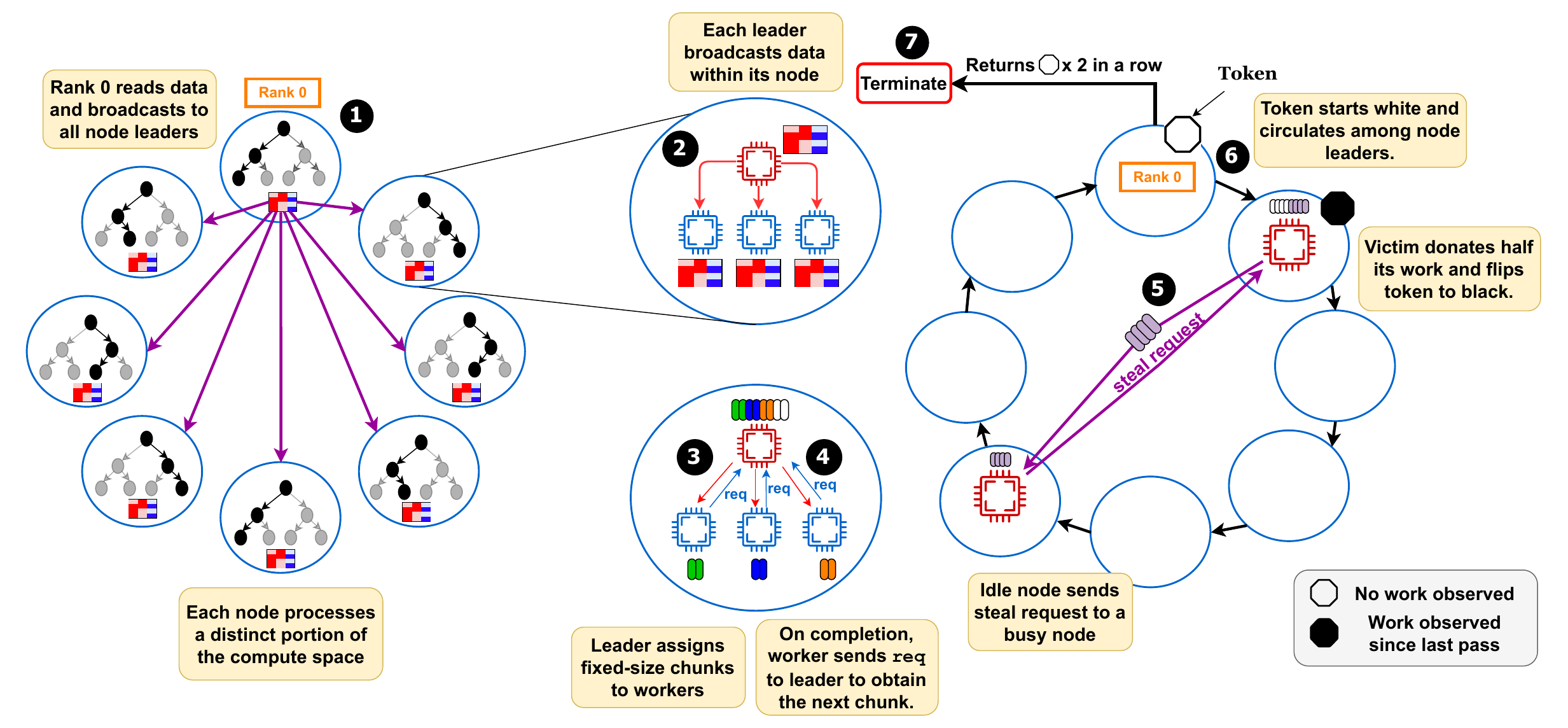}
    \caption{\textbf{Work Distribution Overview}: We use one MPI rank per core and two communicators: a global communicator among node leaders and a local communicator within each node. The local Rank~0 of each node is the node leader and also belongs to the global communicator. \circled{1} Because the database is only a few~MB, Rank~0 reads it from disk and broadcasts the entire database to all node leaders. Each node then computes a distinct subset of $\lambda$ values, i.e., parallelism is in compute. \circled{2} Each node leader broadcasts the data within its node via the local communicator. \circled{3} The leader tracks outstanding $\lambda$ values and hands out fixed-sized work chunks to the workers. \circled{4} When a worker finishes its portion of the compute, it sends a short \texttt{req} to the leader. If work remains, the leader issues another chunk, otherwise the worker becomes idle. \circled{5} If a leader exhausts its local queue, it randomly selects a peer to steal from. A busy victim donates roughly half of its remaining range to the thief, while an idle peer triggers a retry with another node. \circled{6} Termination is decided with a circulating token that starts at Rank~0 in the \emph{white} state (no work observed) and moves around leaders in ring order. Any node that performs work or donates flips the token to \emph{black} (work observed). \circled{7} If the token returns to Rank~0 as black, it is reset to white and circulated again. If it returns white twice in a row, all the work to be computed on has been distributed, and a termination signal is sent.}
    \label{fig:mpi_structure}
\end{figure*}

The workflow of our method consists of several key stages. First, we reorder the binary mutation matrix by gene sparsity, sorting rows from most to least sparse. This sparsity-aware preprocessing enables P-DFS to more effectively prune the search space when identifying candidate multi-hit combinations. To ensure efficient parallelization and workload distribution across MPI ranks, we employ a hierarchical work-stealing strategy (as illustrated in Fig.~\ref{fig:mpi_structure}).
\subsection{Preprocessing}
\label{sec:algo-preproc}

\begin{figure*}[t]
  \centering
  \includegraphics[width=\textwidth]{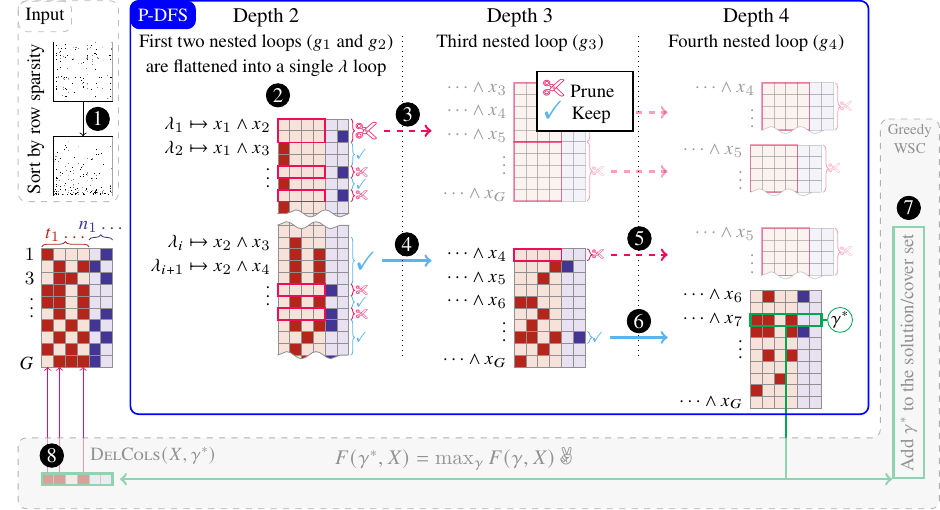}
  \caption{
    \textbf{Pruned depth-first search (P-DFS): pruning algorithm for $h$-hit gene mutations} (with $h=4$):
        In the preprocessing step, the input data is sorted~\circled{1} based on row sparsity to improve pruning efficiency.
    The first two of the four nested loops, with iterators $g_1$ and $g_2$ are flattened into a single $\lambda$-loop~\circled{2} to enable work distribution with a high number of workers.
    In each iteration of the $\lambda$-loop the partial combination using only rows $g_1$ and $g_2$ are computed.
    If all the tumor samples are zero/false, the combination cannot contribute to the set cover (empty cover), nor would the resulting, higher hit combinations, so the entire sub\-tree can be skipped~\circled{3}.
    Otherwise, the algorithm proceeds to the depth~3 nested loop~\circled{4} of the search tree, computing the 3-hit combination (using the 2-hit partial results) and similar eliminates empty covers~\circled{5} or proceeds to depth~4 as needed~\circled{6}.
    The $\gamma^{*}$ combination with the best $F(\gamma, X)$ value is saved~\circled{7} and the samples (columns) covered by it are removed~\circled{8} from the input database for the next iteration.
  }\label{fig:new-workflow}
\end{figure*}

We represent each cohort as a binary mutation matrix with genes as rows and samples as columns (tumor and normal samples are tracked distinctly; see Fig. \ref{fig:new-workflow} for an overview of the workflow). To accelerate P-DFS, we order gene rows from most sparse to least sparse prior to computation. This sparsity-aware ordering helps the search prune aggressively: combinations built from genes that show high sparsity in their mutations can get pruned early, thereby reducing the size of the explored sub-tree. In practice, this simple reordering decreases the number of combination intersections we must evaluate before a branch is pruned.

To make set operations fast, we represent each gene’s samples as a bit-vector and implement primitive operations like intersection, union, and membership count via bitwise operations. This design helps by slashing the memory footprint, hence improving cache residency of hot rows.

% To exploit sparsity at the machine level, we store each row as a bit-vector and implement primitive operations like intersection, union, and membership count via bitwise operations. This design helps by slashing the memory footprint, hence improving cache residency of hot rows. 

\subsection{Core Algorithm}\label{sec:algo-core}

    % \fxnote*{--Emil}{NEED TO REFERENCE Figure~\ref{fig:new-workflow}.}

As discussed in \S\ref{sec:complexity}, the bottleneck of the algorithm is traversing all the $h$-hit gene combinations. Instead of scanning the entire search space, we restructure the traversal where for each $\lambda$, we perform a depth-first expansion with backtracking that prunes as soon as a partial intersection becomes empty. To effectively enable this solution, as mentioned in \S\ref{sec:algo-preproc}, we sort the gene rows by \emph{increasing sparsity} (i.e., from most zero entries to fewer) to trigger empty partial intersections earlier.

For a growing partial set ${g_1,\ldots,g_\ell}$ we maintain the running intersection $y_\ell \leftarrow x_{g_1}\wedge\cdots\wedge x_{g_\ell}$ over the tumor cohort (bitwise \texttt{AND}). 
% If $y_\ell=\mathbf{0}$ (no tumor sample has all $\ell$ mutations) for any $\ell<h$, then every strict superset will also be empty, so the entire subtree can be skipped without evaluation. 
If at any step $y_\ell=\mathbf{0}$ (no tumor sample has all $\ell$ mutations), we immediately backtrack. There is nothing left to gain by adding more genes on that path, so the entire subtree can be skipped without evaluation.
When we reach depth $h$ with $y_h\neq\mathbf{0}$, we score $\gamma$ via the objective $F(\gamma,X)$ and update the current best $(F^*,\gamma^*)$. This simple rule is powerful in sparse data: many branches become empty after just a few \texttt{AND}s, so we examine far fewer than $\binom{G}{h}$ combinations. The P-DFS approach does not change the worst-case time complexity of the original WSC search. In other words, if no branch ever empties, it degenerates to full enumeration of $\binom{G}{h}$ combinations. The algorithm simply aggressively prunes the search space in typical sparse cohorts, so the effective number of candidates evaluated (and the wall time) drops substantially. 

The control flow is made explicit in Algorithm~\ref{alg:nested}: the inner grow-and-check loop maintains the running intersection, and the early-exit test at line~\ref{line:empty} performs the pruning. 

% An interesting structural property of this recursion is that increasing $h$ by one simply adds a level to the same search tree. Therefore, while searching for $N$-hit solutions, every time the recursion reaches depth $N-1$, $N-2$, \emph{etc.}, their feasible partial intersections are already materialized “for free.”

\begin{algorithm}[t]
%\caption{Nested loops\wahib{@Ritvik: use an informative caption}}
  \caption{
    \textbf{P-DFS}: Implementing step~\ref{stp:f} of Algorithm~\ref{alg:greedy} as a depth-first maximum search with backtracking.
  }
\label{alg:nested}
\begin{algorithmic}[1]
\Require $X$ gene mutation matrix, $h$ number of hits
\Ensure $\gamma^{*} = \underset{\gamma \in \Gamma_h}{\operatorname{argmax}} F(\gamma, X)$
\State $F^{*} \gets -\infty$ \Comment{Initialize maximum search}
\For{$g_1 = 1$ \textbf{to} $G - h + 1$} \Comment{Depth 1}
  \State $\gamma \gets {g_1}$ \Comment{The $g_1$-th row of $X$}
  \For{$g_2 = g_1 + 1$ \textbf{to} $G - h + 2$} \Comment{Depth 2}
    \State $\gamma \gets \gamma \cup {g_2}$ \Comment{Implemented as element-wise logical AND}
    \If{$x_{g_1} \land x_{g_2} = \mathbf{0}$}
    \State\textbf{continue}\Comment{Eliminate brach}\label{line:empty}
    \EndIf
    \State \dots
    \For{$g_h = g_{h-1} + 1$ \textbf{to} $G$} \Comment{Depth $h$}
      \State $\gamma \gets \gamma \cup \{g_h\}$
      \State $F \gets F(\gamma, X)$
      \If{$F > F^{*}$} \Comment{Maximum search}
        \State $F^{*} \gets F$; $\gamma^{*} \gets \gamma$
      \EndIf
    \EndFor
    \State \dots
  \EndFor
\EndFor
\State \Return $\gamma^{*}$
\end{algorithmic}
\end{algorithm}

\subsection{Work Distribution: Scaling and Load Balancing}\label{sec:algo-mpi}

% \subsubsection{Exposing Parallelism at Scale}
% Our search space grows combinatorially with the number of genes and the hit depth, so scaling is essential. 
% A natural first attempt is to distribute work by the first gene index only, which yields about 20{,}000 tasks for 20{,}000 genes and caps useful parallelism near 20{,}000 compute units.
% To unlock more concurrency we flatten the nested loops over \(g_1\) and \(g_2\) into a single \(\lambda\) index so each \((g_1,g_2)\) pair becomes an independent task.
% This produces on the order of \(C_2^G \approx \binom{20{,}000}{2}\) tasks and gives every compute unit enough work even at the exascale.

% Irregularity in the pruned search creates a natural load imbalance. Some subtrees collapse after a few intersections while others survive longer and expand. A static assignment will give some workers short subtrees and others long ones, which can produce stragglers.

Our search space grows combinatorially with the number of genes and the hit depth, so we must expose massive concurrency. A naive partition by the first gene yields about 20{,}000 tasks for 20{,}000 genes and caps useful parallelism near 20{,}000 compute units. Instead, we flatten the nested loops over \(g_1\) and \(g_2\) into a single \(\lambda\) index so each \((g_1,g_2)\) pair is an independent task (Fig.~\ref{fig:new-workflow}, \circled{2}), producing \(C_2^G \approx \binom{20{,}000}{2}\) tasks—enough even at the exascale. Furthermore, pruning makes traversal cost per task highly \emph{irregular}: some subtrees collapse after only a few intersections while others survive and expand. Equal counts of \(\lambda\)s therefore do not imply equal work, and a static partition risks idle ranks and stragglers. This motivates the MPI-based, dynamically assigned work distribution described below (see the control-flow overview in Fig.~\ref{fig:mpi_structure}).

\paragraph{MPI Execution Model and Data Residency}
% \fxnote*{Original in the comments.}
We scale with MPI, running one rank per physical core, and assign disjoint $\lambda$ sub-intervals to each rank for independent computation. Given that the dataset is just a few megabytes in size, each rank holds the entire input table. Furthermore, each rank obtains a sub\-interval of \(\lambda\)s from \([1, \ldots, C_2^G]\).
By decoding \(\lambda\) back to \((g_1, g_2)\), the state of the program corresponds to step~\ref{line:empty} in Algorithm~\ref{alg:nested}, and it can proceed to the first level of pruning if no samples share the mutations of both genes.
% Each rank holds a private copy of the Gene–Sample table in compact \texttt{uint64\_t} bitsets.
% Combinations are generated on the fly by decoding \(\lambda \rightarrow (g_1, g_2)\) and intersecting the two rows locally, which avoids enumerating the full set of possible combinations and hence avoids saturating the local memory
% - \ritvik{May need to rephrase.}

\paragraph{Baseline: centralized master–worker}
We begin with a master–worker design in which rank~0 is the master and all other ranks are workers. The master maintains a single global pool of \(\lambda\) indices and dispenses fixed-size chunks with indices \(\langle \texttt{start}, \texttt{end} \rangle\) on demand. Each worker stores only the bounds of its current chunk, computes directly over its private copy of the table, and, upon completion, sends a short request to the master for the next chunk. Using a fixed chunk size keeps assignment constant-time and minimizes bookkeeping, so control traffic remains light. In other words, workers communicate only when pulling new work and when receiving the final termination signal from the master.

\paragraph{Limitations of centralized master-worker}
A single global master would saturate on large machines. Every request would target one rank, queues would grow, and the MPI progress engine would spend time on control traffic rather than computation. Prior studies of dynamic scheduling report that single point task coordination limits scalability and motivates multi-level coordinators\cite{yang_large-scale_2013, perarnau_victim_2014}. We therefore move to a hierarchical organization that respects the machine topology. 

\paragraph{Topology-aware hierarchical organization}
All ranks on a node form a local communicator and designate rank~0 as the node leader. All node leaders form a second communicator that spans nodes. Leaders manage work inside the node, and coordinate only coarse actions across nodes, which reduces cross-node traffic and removes the single choke point.

\paragraph{Two-level work stealing}
Work stealing operates at two levels. Inside a node, workers never steal from each other. They always ask the node leader for the next interval. Across nodes, leaders both request and donate work. When a leader runs out of local work it selects a peer leader and issues a steal request. Peer selection is randomized with a small number of retries in case the chosen leader has no more work to give. This spreads requests across the ring of leaders and avoids hot spots. When a leader receives a steal request it answers from its own remaining interval. The leader replies with the upper half of its remaining \(\lambda\)-interval. This splits its workload roughly in two, hands a single contiguous block, which is essentially a \(\langle \texttt{start}, \texttt{end} \rangle\) interval, to the thief. The victim also records that a donation occurred via a local flag, which is used later during coordination.

\paragraph{Barrier-free termination detection}
Termination and global progress use a token on the leader communicator. Leaders form a logical ring. The root creates a white token and sends it to the next leader. A leader forwards the token only when it is idle and has no local work, and if it has donated work since it last saw the token it turns the token black before forwarding. If the root receives a white token twice in succession while it is idle, there is no outstanding work. The root then sets the termination flag via MPI one-sided communication, issuing an \texttt{MPI\_Put} into a shared RMA window, and each leader notifies its workers to exit their request loop. This protocol detects quiescence without a global barrier and remains robust under random steals, since the black color records backward donations and suppresses premature termination.

\paragraph{Hierarchical collectives: Broadcast and AllReduce}
Reductions and broadcasts follow an identical hierarchy structure. First, within each node, processes reduce onto a designated leader using intranode collectives that exploit shared memory. The leaders then perform the internode collective on the leader communicator. Finally, each leader distributes the result back to its local workers via an intranode broadcast. Although this hierarchical framework increases memory consumption, it reduces message traffic on the network fabric and shortens the critical path for large collectives, which is crucial on systems where MPI buffers and progress threads face significant contention at scale (realistic benchmarks demonstrate up to a 7× speedup for all-reduce using these two-level structures compared to standard MPI\_Allreduce).
% \vspace{15px}
\paragraph{Communication profile and the MPI-only choice}
Given that we replicate the gene–sample mutation table per rank, all the inner loop work memory stays local to the node and communication is confined to work requests, inter-leader steals, hierarchical collectives, and a final termination flag. We prefer rank-per-core over hybrid MPI+OpenMP for this irregular, fine-grained workload because thread schedulers introduce overheads from frequent returns to the scheduler, extra barriers, and affinity or NUMA pitfalls, which can make hybrid slower than pure MPI without careful tuning \cite{diener_improving_2017, cappello_mpi_2000}. 

\subsection{Pre-sorted DFS Finds Smaller Covers at Higher Hit Thresholds}
\label{subsec:k4_analysis}

% Requires: \usepackage{multirow}
\begin{table}[t]
    \centering
  \renewcommand{\arraystretch}{1.3}
  \vspace{5px}
    \caption{Comparison of Proposed Algorithm vs. Original WSC Solution Size for Various Cancers and Hit Sizes}
    \begin{tabular}{lccc}
        \toprule
        \textbf{Cancer} & \textbf{Hit Size } & \textbf{Proposed Algorithm} & \textbf{Original WSC} \\ 
        \midrule
        BLCA & 2 & 17 & 15 \\ 
        GBM & 2 & 11 & N/A \\ 
        OV & 2 & 9 & N/A \\ 
        UCEC & 2 & 5 & 9 \\ 
        \hdashline
        BLCA & 3 & 16 & 12 \\ 
        BRCA & 3 & 12 & 8 \\ 
        COAD & 3 & 12 & 10 \\ 
        KIRC & 3 & 9 & 6 \\ 
        LUAD & 3 & 13 & 10 \\ 
        SKCM & 3 & 11 & N/A \\ 
        \hdashline
        BLCA & 4 & 16 & 19 \\ 
        HNSC & 4 & 13 & 17 \\ 
        \bottomrule
    \end{tabular}
    \label{tab:solution_sizes}
\end{table}
%An advantage of our algorithm is that we find better solutions, i.e. smaller set covers, at higher hits (see Table~\ref{tab:solution_sizes}). Note that since the solution space increase exponentially with number of hits, getting smaller set covers in higher number of hits can dramatically improve the performance since we terminate earlier when we since a smaller cover set. In this section we discuss the reason for why our algorithm enables smaller cover sets at higher number of hits.
An advantage of our algorithm is that it yields better solutions, i.e., smaller set covers, at higher hit thresholds. As shown in Table~\ref{tab:solution_sizes}, with our algorithm we reduce the solution size at four hits by 80\% average.
Since the size of the solution space grows exponentially with the number of hits, obtaining smaller cover sets at higher number of hits provides a substantial performance benefit, as the search can terminate earlier once a minimal cover is found. In this section, we analyze why the proposed algorithm tends to produce smaller cover sets at higher hit thresholds.

As introduced in Eq.~(\ref{eq:goal}), our goal is to find the set of combinations that has a minimal combined weight. While the baseline explores combinations randomly, the proposed algorithm sorts genes, i.e. matrix rows, by sparsity (from fewer to more mutations), and uses P-DFS to prune the search space of combinations.

\subsubsection{Effect of sparsity at high $h$}
When the number of hits $h$ is small (e.g., $h=2$), dense genes with many mutations quickly increase the hit coverage $\mathcal{C}$ defined in Eq.~(\ref{eq:cover-gamma}) even if they overlap strongly.
At higher $h$, however, overlapping dense genes provide less new information,
because many samples already share the same mutations.
Thus, the expected marginal gain of a gene $g$ given the current partially constructed combination of genes $\gamma$, $\Delta_h(g \mid \gamma) = \mathcal{C}(\gamma \cup \{g\}) - \mathcal{C}(\gamma)$, decreases more sharply for dense genes as $h$ increases.
Sparse genes, which cover fewer but more distinct samples, maintain higher marginal gain when $h$ is large.
\subsubsection{Pruning advantage}
Because our search visits sparse genes first, partial solutions grow coverage efficiently with minimal redundancy.
This makes the upper bound on remaining possible gain, $U_h(\gamma) = \max_{S \subseteq \Gamma_h \setminus \gamma} \mathcal{C}(\gamma \cup S) - \mathcal{C}(\gamma)$, drop faster as the search deepens.
Branches that cannot reach the target are pruned earlier.
At higher $h$, the redundancy among dense genes strengthens this effect,
reducing the effective branching factor and shrinking the explored tree from $O(2^n)$ toward $O(2^L)$,
where $L$ is the number of sparse, high-value genes.

\subsubsection{Resulting behavior}
This explains why the proposed method produces smaller covers at $h=4$.
For large $h$, sparse-first ordering better matches the structure of the residual samples,
yields higher marginal gains per added gene, and enables tighter pruning.
In contrast, random or dense-first exploration revisits many redundant regions,
leading to larger covers or slower convergence.
Empirically, this aligns with our results, where P-DFS achieves
smaller solution sizes at 4-hit thresholds.

\section{Results and Discussion}
\label{sec:results}

\subsection{Experimental Platform: Fugaku}
\label{sec:fugaku}

% \wahib{@Ritvik: this should move to the results section}

All experiments were conducted on the supercomputer Fugaku, located in Kobe, Japan. Each node contains a single Fujitsu A64FX processor with 48 compute cores and 32~GB of HBM2 memory. We use Fujitsu MPI and run one MPI rank per core with no intra-rank threading. This choice avoids thread-scheduler overheads and lets each rank maintain a private copy of the data. Our datasets are only a few megabytes per cohort, so replication across 48 ranks comfortably fits within the 32~GB node memory budget and removes contention on shared data structures.

Nodes are connected by the Tofu-D interconnect in a six-dimensional mesh–torus topology. We request compact allocations that form near-cubic sub-tori so that the job spans similar extents in each of the six dimensions. This placement reduces average hop distance, preserves bisection bandwidth, and helps keep collective and point-to-point costs stable as we scale out. The node counts in our strong-scaling studies \{192, 384, 768, 1{,}536, 3{,}072\} were chosen because they tile the 6D topology cleanly and allow the scheduler to pack jobs contiguously.

Our MPI configuration follows the algorithmic design. A single leader rank per job performs interval assignment, and all other ranks act as workers that compute against their private tables. We bind one rank to one physical core and pin memory locally to improve cache residency on A64FX. Communication consists of short messages to pull work and a few global notifications at the end of the run. On Tofu-D these messages traverse few hops under compact placement, which keeps the communication term nearly constant across scales, as observed in \S\ref{sec:results-strong-scaling}. The combination of rank-per-core, data replication, and compact topology-aware placement is what allows the pruned workload to remain compute dominated over a wide range of node counts.

\subsection{Dataset}
\label{sec:dataset}
Somatic mutation data in the Mutation Annotation Format (MAF) was acquired from The Cancer Genome Atlas (TCGA)~\cite{noauthor_cancer_2022}. All mutations were identified via the Mutect2 software, a commonly utilized tool in the genomic community. Each cancer type was processed individually, extracting gene-level mutations paired with patient sample identifiers. Silent mutations were excluded to focus on potentially functional variants. Following preprocessing, each dataset was divided into train and test sets at 75\% and 25\%, respectively.

\subsection{Pruning Efficiency and End-to-End Runtime}
\label{sec:results-pruning-runtime}

A central outcome of our approach is that we only visit a tiny fraction of the potential $h$-hit combinations. For each cohort, we tracked the exact number of combinations that the P-DFS traversal evaluated and compared it to a non-pruning baseline whose search space equals the combinatorial maximum $\binom{20{,}000}{h}$. Cohorts are paired with the hit sizes reported in~\cite{anandakrishnan2019estimating}, %the study of Anandakrishnan \emph{et al.}~\cite{anandakrishnan2019estimating}, 
which estimates the number of hits for 17 cancer types with at least 200 samples. 

\begin{figure}[t]
    \centering
    \includegraphics[width=\linewidth]{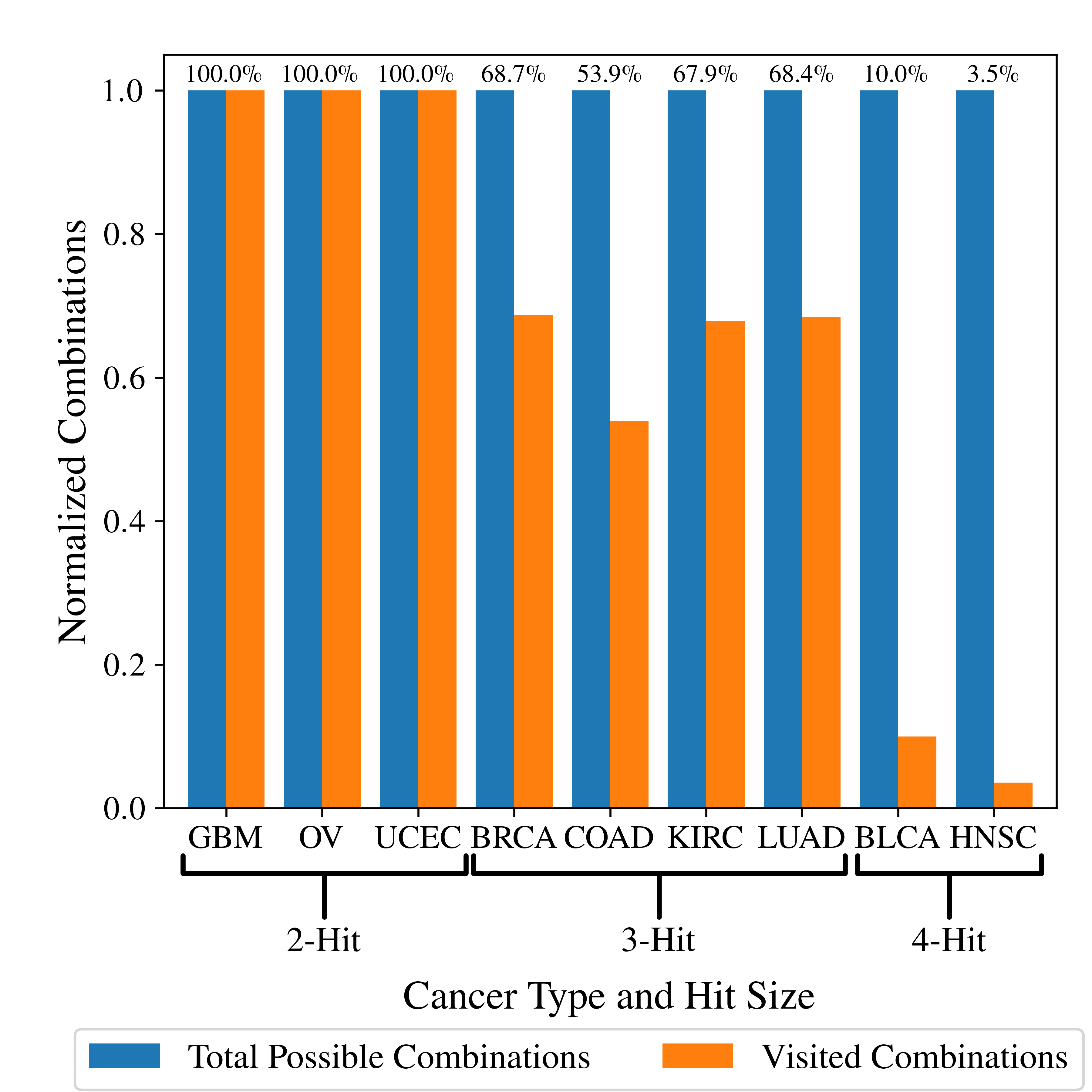}
    \caption{Pruned search vs.\ theoretical maximum. For each cohort and hit size, the total combinations $\binom{20{,}000}{h}$ are normalized to $1.0$ (blue). The orange bars show the fraction of combinations actually visited by our sparsity-driven P-DFS method.}
    \label{fig:visitation_plot}
\end{figure}

Fig.~\ref{fig:visitation_plot} reports, for multiple cohorts and hit sizes, the normalized number of combinations that are actually visited by our algorithm. The blue bars represent the total search space and are normalized to $1.0$ by construction. The orange bars show the fraction we visit. Lower orange bars indicate more pruning. We observe a consistent drop in visitation as $h$ increases. This behavior is expected because early intersections become empty much more often at higher $h$. In sparse mutation matrices, the probability that a sample carries all genes in a growing partial set decreases rapidly with each added gene. The result is aggressive pruning at higher hit sizes and a dramatic reduction in evaluated candidates.

Pruning translates directly into wall-time gains. We study this effect with a 4-hit run for BLCA across 192, 384, 768, 1{,}536, and 3{,}072 nodes. We also plot an ideal strong-scaling line for the pruned workload to check whether runtime falls in inverse proportion to node count. For a fixed workload of size $MN_t$ the ideal time on $N$ nodes is $T_{\text{ideal}}(N)=\frac{c\,MN_t}{N}$, where $M$ is the number of combinations actually visited by the algorithm, $N_t$ is the number of tumor samples, and $c$ is a platform constant that captures per-combination cost. We calibrate $c$ from the 192-node run, $c=\frac{T_{\text{measured}}(192)\cdot 192}{MN_t}$, which yields $T_{\text{ideal}}(N)=T_{\text{measured}}(192)\cdot\frac{192}{N}$.

The bounded curve lies close to this line up to 768 nodes but shows slight divergence beyond 768 nodes. 

To highlight the value of pruning, we also executed a baseline run at 3{,}072 nodes.
% These node counts double at each step and they map cleanly onto the TOFU-D 6D mesh interconnect of Fugaku. 
This job did not finish within the four-hour allocation. Based on the completed portion, we estimate a time to solution of about 52 hours on 3{,}072 nodes. This projected time is much slower than the P-DFS method even at 192 nodes. The gap reflects the massive volume of candidates that fail the first few intersection checks and yet must still be evaluated when pruning is disabled.

\begin{figure}[t]
    \centering
    \includegraphics[width=0.98\linewidth]{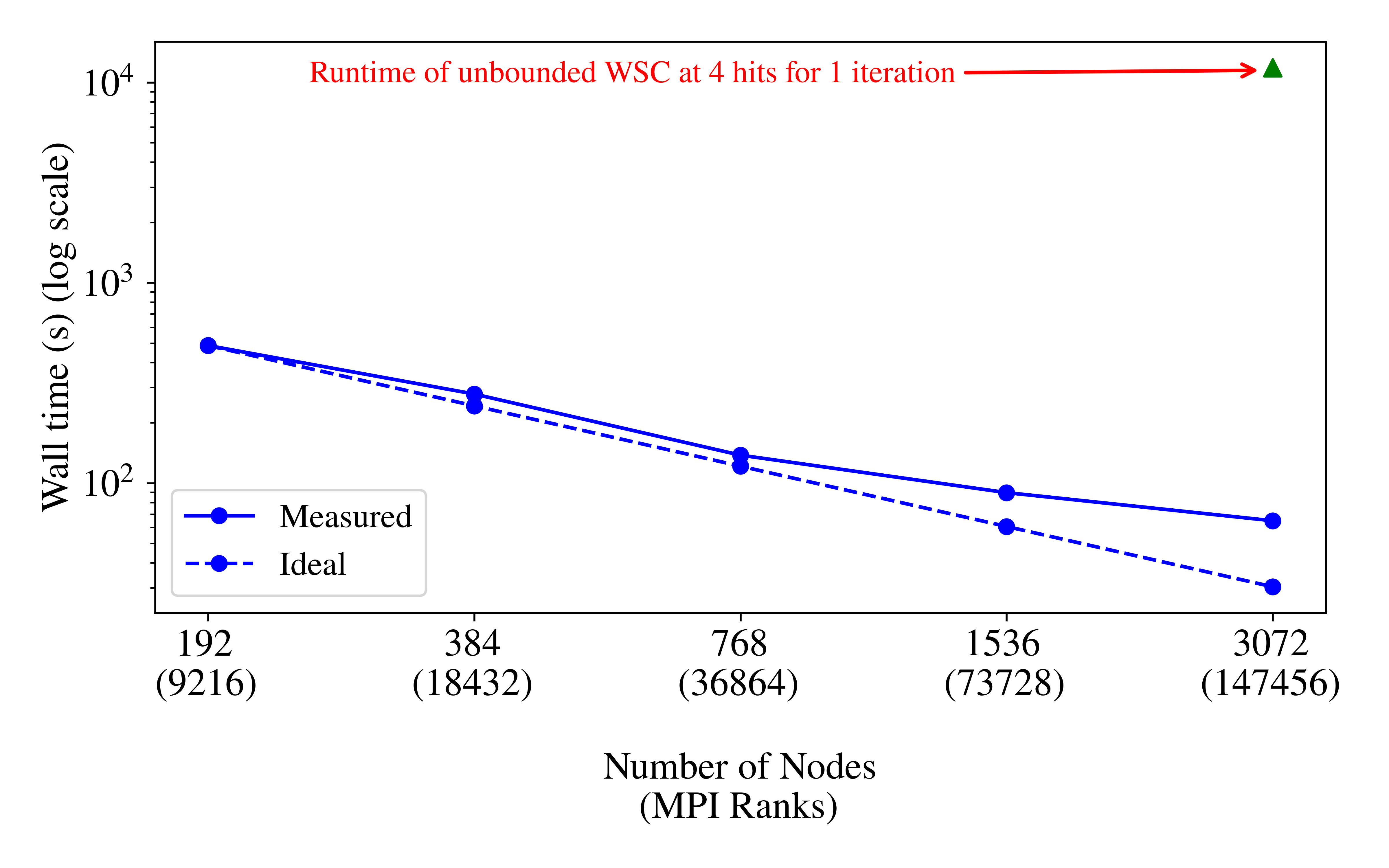}
    \caption{\textbf{Wall time vs.\ scale for 4 hits}. Measured wall times for the P-DFS runs across nodes, together with the ideal strong-scaling line.}
    \label{fig:wall_time_ub_vs_b}
\end{figure}

Taken together, Figs.~\ref{fig:visitation_plot} and~\ref{fig:wall_time_ub_vs_b} show that our design delivers much better performance compared to the brute-force weighted set cover algorithm. The algorithm prunes vast regions of the search space before computing the weights.

\subsection{Accuracy on Held-out Data}
\label{sec:results-accuracy}

We evaluate accuracy of the P-DFS method with a random $75\%$–$25\%$ train–test split per cohort. On the training set, the algorithm learns a small set of $h$-hit gene combinations using P-DFS weighted set cover method. We then freeze those combinations and assess coverage on the disjoint Test set. For each cohort we report:
\begin{itemize}
    \item \textbf{Tumor Covered}: number of tumor samples that contain at least one learned $h$-hit combination.
    \item \textbf{Normal Covered}: number of normal samples that contain at least one learned $h$-hit combination.
\end{itemize}

\paragraph{Training performance}\label{sec:training-perf}
Table~\ref{tab:training_results} summarizes results on the training split (TC, TT, NC, NT, Sens.\ and Spec.\ correspond to number of covered tumor samples, total number of tumor samples, number of covered normal samples, total number of normal samples, sensitivity and specificity, respectively). Sensitivity is $1.0$ for all cohorts, and specificity is very high, with small leakage into normals for BRCA, COAD, KIRC, OV, and UCEC. These results are expected because the model is optimized on the training set. Furthermore, these values represent an upper bound on achievable performance for a given $h$.

\begin{table}[t]
  \vspace*{15pt}
  \renewcommand{\arraystretch}{1.3}
  \caption{
    Training split ($75\%$). Tumor and normal coverage and derived sensitivity/specificity for learned combinations.
  }\label{tab:training_results}
  \centering
  %%% \resizebox{\columnwidth}{!}{%
  \begin{tabular}{lccccccc}
    \toprule
    %%% Cancer & Hits & Tumor Covered & Tumor Total & Normal Covered & Normal Total & Sensitivity & Specificity \\ \hline
    Cancer & Hits & TC & TT & NC & NT & Sens. & Spec. \\
    \midrule
    BLCA & 4  & 309 & 309 & 0 & 248 & 1.0 & 1.0    \\
    HNSC & 4  & 382 & 382 & 0 & 248 & 1.0 & 1.0    \\
    BLCA & 3  & 309 & 309 & 0 & 248 & 1.0 & 1.0    \\
    BRCA & 3  & 786 & 786 & 8 & 248 & 1.0 & 0.9677 \\
    COAD & 3  & 325 & 325 & 4 & 248 & 1.0 & 0.9839 \\
    KIRC & 3  & 288 & 288 & 1 & 248 & 1.0 & 0.9960 \\
    LUAD & 3  & 426 & 426 & 0 & 248 & 1.0 & 1.0    \\
    SKCM & 3  & 352 & 352 & 0 & 248 & 1.0 & 1.0    \\
    GBM  & 2  & 297 & 297 & 0 & 248 & 1.0 & 1.0    \\
    OV   & 2  & 319 & 319 & 2 & 248 & 1.0 & 0.9919 \\
    UCEC & 2  & 406 & 406 & 8 & 248 & 1.0 & 0.9677 \\
    BLCA & 2 & 309 & 309 & 0 & 248 & 1.0 & 1.0     \\
    \bottomrule
    \end{tabular}
    %%% }
\end{table}

\paragraph{Generalization}
Table~\ref{tab:test_results} reports performance on the disjoint $25\%$ test split using the combinations learned on training (column names detailed in \S\ref{sec:training-perf}). Sensitivity remains high across cohorts ($0.85$–$0.98$), and specificity is similarly strong ($0.81$–$0.99$). The small reduction relative to training reflects that some held-out tumors lack the exact $h$-hit patterns selected during training.

\begin{table}[t]
  \renewcommand{\arraystretch}{1.3}
  \caption{
    Held-out Test split ($25\%$). Coverage-based sensitivity and specificity using combinations learned on the Training split.
  }\label{tab:test_results}
  \centering
  %%% \resizebox{\columnwidth}{!}{%
  \begin{tabular}{lccccccc}
    \toprule
    % Cancer & Hits & Tumor Covered & Tumor Total & Normal Covered & Normal Total & Sensitivity & Specificity \\
    Cancer & Hits & TC & TT & NC & NT & Sens. & Spec. \\
    \midrule
    BLCA & 4 &  88 & 103 & 16 & 83 & 0.8544 & 0.8072 \\
    HNSC & 4 & 113 & 128 & 14 & 83 & 0.8828 & 0.8313 \\
    BLCA & 3 &  89 & 103 & 12 & 83 & 0.8641 & 0.8554 \\
    BRCA & 3 & 257 & 262 & 13 & 83 & 0.9809 & 0.8434 \\
    COAD & 3 &  97 & 109 &  8 & 83 & 0.8899 & 0.9036 \\
    KIRC & 3 &  92 &  97 &  8 & 83 & 0.9485 & 0.9036 \\
    LUAD & 3 & 124 & 143 & 14 & 83 & 0.8671 & 0.8313 \\
    SKCM & 3 & 106 & 118 &  7 & 83 & 0.8983 & 0.9157 \\
    BLCA & 2 &  88 & 103 &  6 & 83 & 0.8544 & 0.9277 \\
    GBM  & 2 &  95 &  99 &  1 & 83 & 0.9596 & 0.9880 \\
    OV   & 2 & 102 & 107 &  3 & 83 & 0.9533 & 0.9639 \\
    UCEC & 2 & 131 & 136 &  5 & 83 & 0.9632 & 0.9398 \\
    \bottomrule
  \end{tabular}
  %%% }
\end{table}

\paragraph{Takeaways}
Across ten cohorts and hit sizes guided by \cite{anandakrishnan2019estimating}, our approach achieves near-perfect sensitivity and very high specificity on training, and maintains strong accuracy on test data. High test sensitivity indicates that the selected $h$-hit patterns capture tumor signal that generalizes to unseen patients. High test specificity shows that normal tissue rarely matches these patterns. Combined with the pruning and scaling results, these accuracy outcomes demonstrate that the algorithm is biologically discriminative at scale.

\subsection{Load Balancing and the Impact of Work Stealing}
\label{sec:results-load-balance}

% Irregularity in the pruned search creates natural load imbalance. Some subtrees collapse after a few intersections while others survive longer and expand. A static assignment will give some workers short subtrees and others long ones, which could produce lots of stragglers.
\textbf{We quantify the effect of irregularity by measuring the worker running time on BLCA with 4 hits at 1{,}536 nodes.} Worker running time records only periods when a worker is computing, not idle or waiting on the leader. We run the same configuration with work stealing enabled and disabled and collect per-worker histograms.

\begin{figure}[t]
    \centering
    \includegraphics[width=\linewidth]{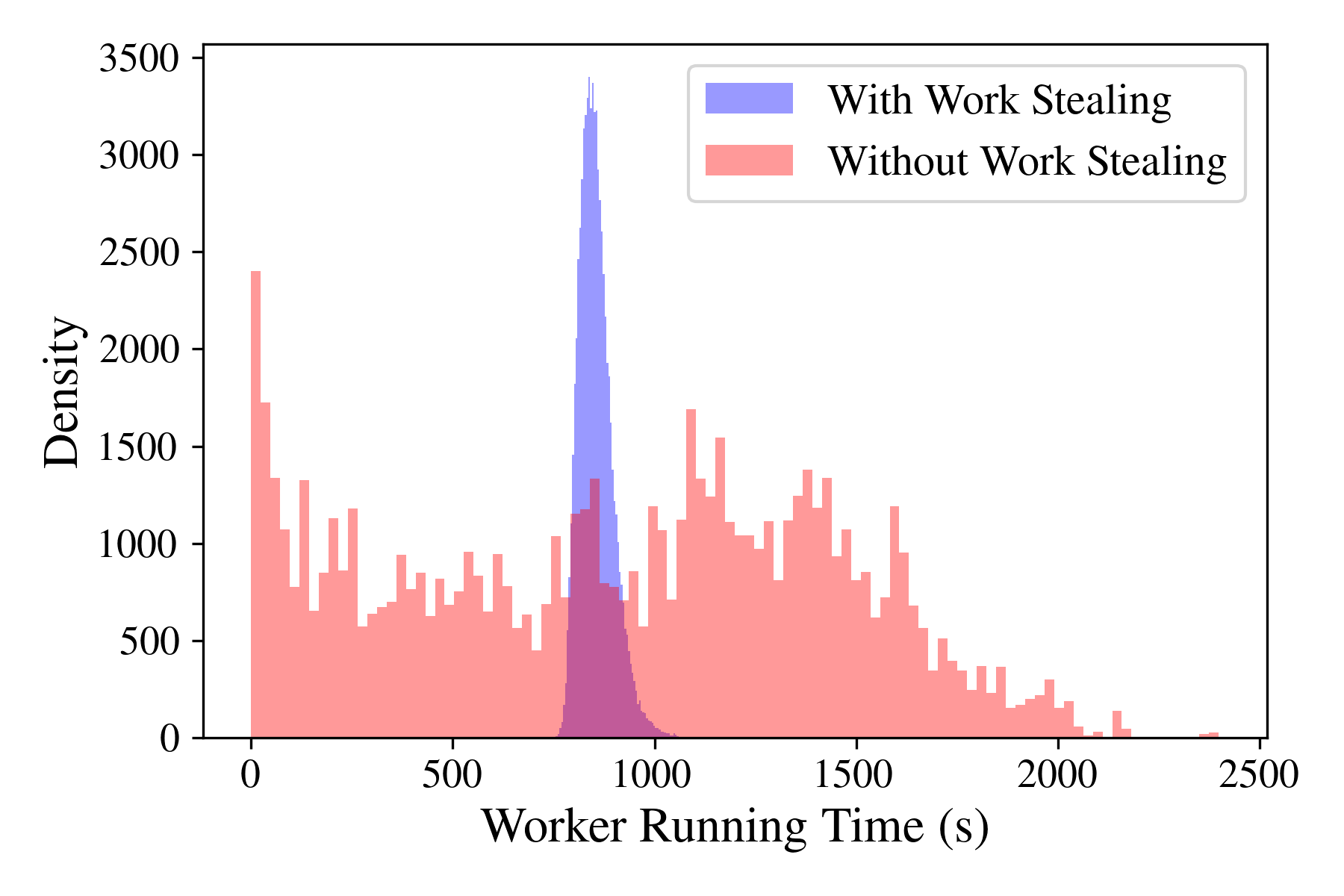}
    \caption{Per-worker running-time distributions for BLCA, 4 hits, 1{,}536 nodes.}
    \label{fig:work_distribution}
\end{figure}

Fig.~\ref{fig:work_distribution} shows the distributions. Without stealing the distribution is broad with heavy right tails, which indicates that many workers receive subtrees that are much larger than average. With stealing the distribution collapses into a sharp peak. The standard deviation drops from $544.186$ seconds without stealing to $41.9599$ seconds with stealing, a $13\times$ reduction in spread. Likewise, \textbf{we observed average worker idleness likewise falls from $63.61\%$ (no stealing) to $22.31\%$ (with stealing), reinforcing the benefits of work stealing.} This indicates that workers finish at nearly the same time when stealing is enabled. The reduced spread and average idleness of the workers aligns with the lower wall times reported in Fig.~\ref{fig:wall_time_ub_vs_b}, since long stragglers can dominate end-to-end time.

\subsection{Strong Scaling Across Hit Sizes}
\label{sec:results-strong-scaling}

We now quantify the parallel efficiency across hit sizes using speedup from a 192-node baseline. For each cohort we keep the algorithm and implementation fixed: P-DFS method with work stealing, and only vary the node count $N\in\{192,384,768,1536,3072\}$. A key trend stands out: scaling improves as the hit count increases, with BLCA (4 hits) tracking the ideal most closely, BRCA (3 hits) in the middle, and OV (2 hits) saturating early. Before interpreting the speedup curves, we first examine what fraction of wall time is computation versus communication.

\begin{figure}[t]
    \centering
    \includegraphics[width=\linewidth]{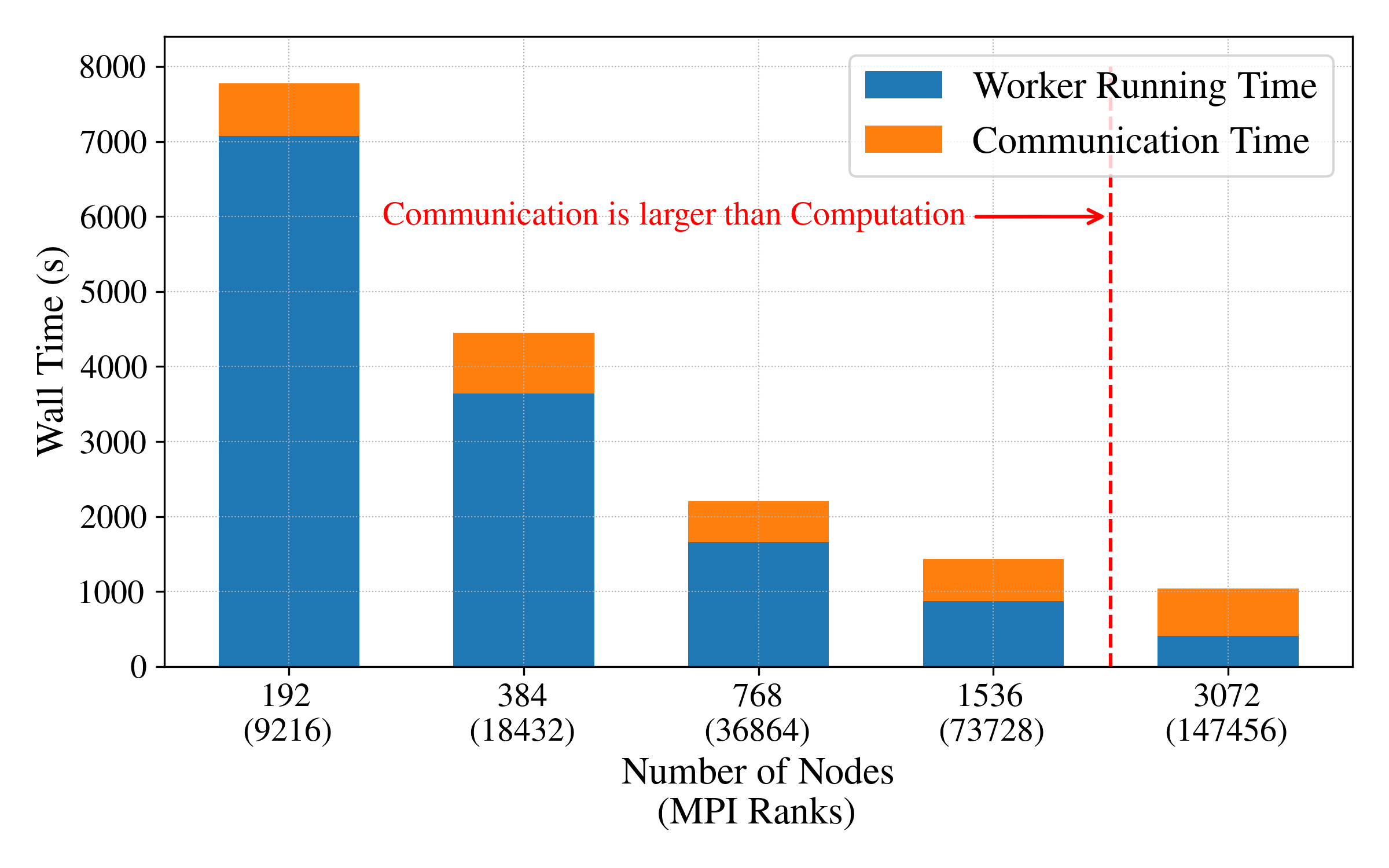}
    \caption{Wall-time breakdown of BLCA at 4 hits.}
    \label{fig:time_distribution}
\end{figure}

Fig.~\ref{fig:time_distribution} decomposes wall time into worker running time and communication for BLCA at 4 hits. Communication is comparatively constant across scales, while worker running time falls with node count as expected for a compute-dominated phase. When the compute term is large the constant is amortized and strong scaling holds. When the compute term is small the communication limits the additional speedup.

With this context we return to the speedup curves. After pruning, the useful work is proportional to $MN_t$, where $M$ is the number of combinations actually visited and $N_t$ is the number of tumor samples. A simple model $T(N) \approx \frac{c\,MN_t}{N} + T_{\text{comm}}$ explains the observations. For low hit counts, $M$ is small because intersections go empty early, so $\frac{c\,MN_t}{N}$ becomes comparable to the constant $T_{\text{comm}}$ at modest $N$. The OV (2-hit) curve therefore plateaus and even dips as noise and fixed overheads dominate. As the hit count increases, more combinations survive pruning, so $M$ is larger and the compute term dominates over a wider range of $N$. BRCA (3 hits) improves over OV, and BLCA (4 hits) stays close to the ideal line because the run remains compute bound even at 3{,}072 nodes. 

\begin{figure}[t]
    \centering
    \includegraphics[width=\linewidth]{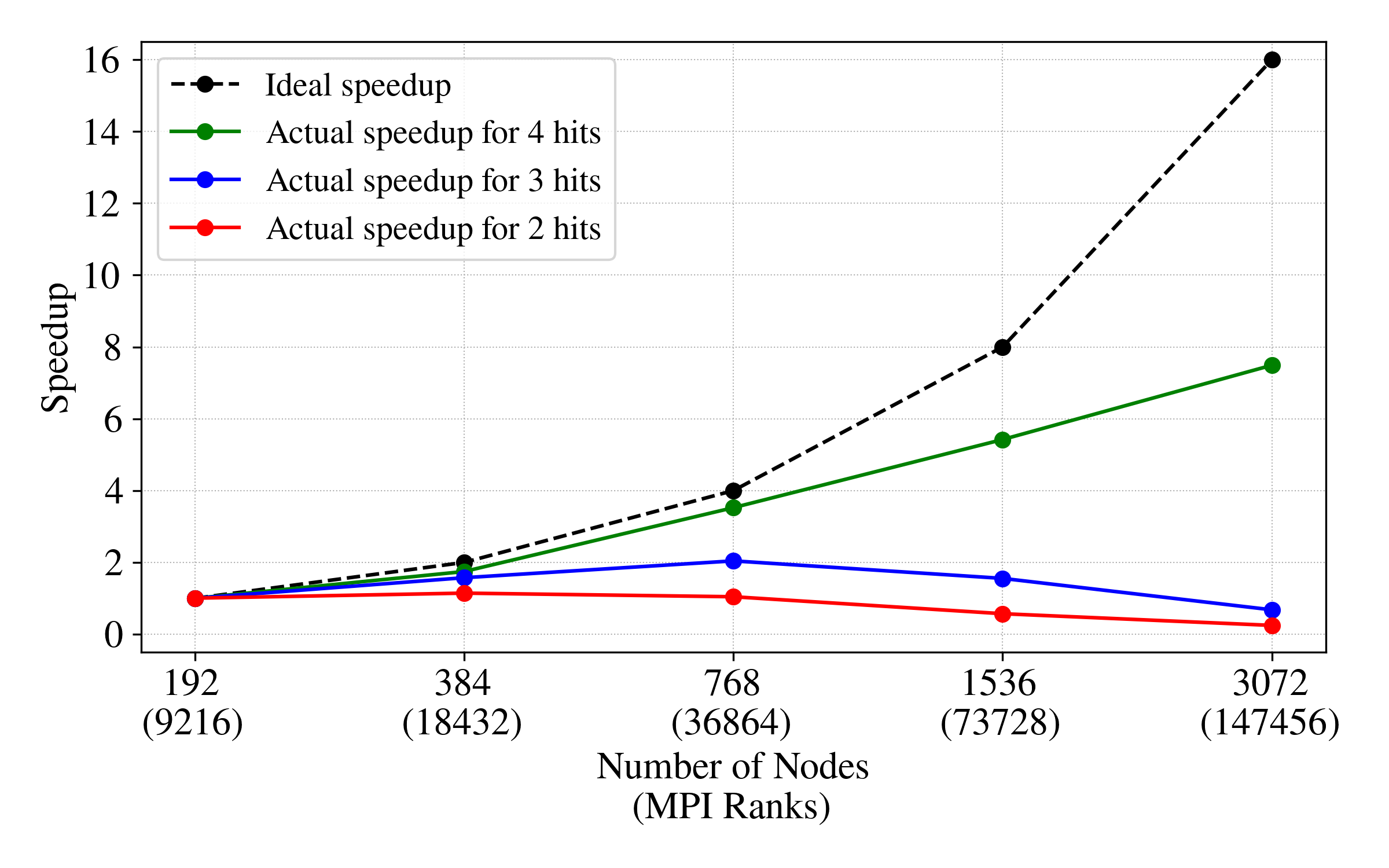}
    \caption{Strong scaling speedup from 192 to 3{,}072 nodes for BLCA 4-hit, BRCA 3-hit, and OV 2-hit datasets.}
    \label{fig:strong_scaling}
\end{figure}

\section{Conclusion}
\label{sec:conclusion}
We present a sparsity-aware depth-first search algorithm with backtracking 
% variant of 
\review{that serves as our main contribution: a pruning technique for}
the weighted set cover (WSC) 
% algorithm
\review{pipeline}, called \emph{pruned depth-first search}, for discovering multi-hit gene combinations. By sorting genes by sparsity, maintaining running bitset intersections, and backtracking as soon as partial covers are found to be empty, the search algorithm discards vast swaths of the combinatorial space \review{(i.e., reduces the candidate search space)} before scoring \review{with the (standard) WSC selection step}. To expose parallelism, we flatten the first two nested loops of the WSC algorithm into a single $\lambda$ index and distribute contiguous $\langle \texttt{start},\texttt{end}\rangle$ intervals, enabling high worker counts. A hierarchical MPI design, with per-node leader, lightweight request/response assignment, and barrier-free termination, keeps control traffic low and avoids single-master bottlenecks.

Empirically, pruning grows more effectively with the hit threshold. In other words, higher $h$ induces emptier partial intersections, shrinking the visited space and reducing wall time relative to exhaustive WSC enumeration. On Fugaku, we observe that strong scaling improved as we increase the number of hits being computed. Furthermore, we observe tighter load-balance distributions when hierarchical scheduling is enabled. Across cohorts, the resulting combinations achieve high sensitivity and specificity on held-out data (via random $75\%$–$25\%$ splits), indicating that sparsity-driven pruning, paired with weighted set-cover scoring yields patterns that generalize. Taken together, these results demonstrate that P-DFS \review{pruning} converts an otherwise intractable enumeration into a high-throughput search at scale.  

\section*{Code Availability}
The reference implementation and scripts to reproduce our results are available at
\url{https://github.com/RitvikPrabhu/P-DFS-Multihit-WSC}.

\section*{Acknowledgments}
We thank Niteya Shah for constructive comments and feedback on an early draft of this paper.

\bibliographystyle{IEEEtran}
\bibliography{references, ritvik-zotero}%, bernie_references}

% {\color{red} DON'T FORGET}
\clearpage
% \balance
% \appendix
% \include{ipdps26_ad_ae}
\clearpage

\twocolumn[%
{\begin{center}
\Huge
Appendix: Artifact Description        
\end{center}}
]

%%%%%%%%%%%%%%%%%%%%%%%%%%%%%%%%%%%%%%%%%%%%%%%%%%%%%
%  AD Appendix
%%%%%%%%%%%%%%%%%%%%%%%%%%%%%%%%%%%%%%%%%%%%%%%%%%%%%

\appendixAD

\section{Overview of Contributions and Artifacts}

\subsection{Paper's Main Contributions}

\begin{description}
\item[$C_1$] We introduce Pruned Depth-First Search (P-DFS), which is a DFS/backtracking pruning method that exploits the high sparsity of tumor mutation matrices to prune infeasible $h$-hit gene subsets early, drastically reducing the effective combinatorial search space before Weighted Set Cover (WSC) scoring/selection.
\item[$C_2$] We integrate P-DFS with HPC-focused optimizations, including bitset-based set operations (via bitwise primitives) and a distributed, topology-aware work distribution strategy (hierarchical collectives and work stealing) designed to execute the pruned search efficiently at scale. 
\item[$C_3$] We evaluate pruning effectiveness, solution/coverage behavior, and strong-scaling performance on real cancer datasets, reporting large reductions in explored combinations for 4-hits and substantial end-to-end speedups over exhaustive enumeration, including results measured at very large rank counts on Fugaku.
\end{description}

% \artexpl{
% Provide a list of all main contributions of the paper.
% }

% \artsampl{

% }

\subsection{Computational Artifacts}

\begin{description}
\item[$A_1$] \textbf{Core implementation} (DOI: \texttt{N/A}).
Code repository: https://github.com/RitvikPrabhu/P-DFS-Multihit-WSC
\end{description}

% \artexpl{
% List the computational artifacts related to this paper along with their respective DOIs. Note that all computational artifacts may be archived under a single DOI.
% }

% \artsampl{
% \begin{description}
% \item[$A_1$] https://doi.org/YY.YYYY/zenodo.0XXXXX
% \item[$A_2$] https://doi.org/ZZ.YYYY/zenodo.1XXXXX
% \item[$A_3$] https://doi.org/ZZ.YYYY/zenodo.2XXXXX
% \end{description}
% }

% \artexpl{
% Provide a table with the relevant computational artifacts, 
% highlight their relation to the contributions (from above) and 
% point to the elements in the paper that are reproducible by each artifact, e.g., 
% which figures or tables were generated with the artifact.
% }

% \artsampl{
% \begin{center}
% \begin{tabular}{rll}
% \toprule
% Artifact ID  &  Contributions &  Related \\
%              &  Supported     &  Paper Elements \\
% \midrule
% $A_1$   &  $C_1$ & Tables 1-2 \\
%         &        & Figure 3\\
% \midrule
% $A_2$   &  $C_2$ & Tables 2-3 \\
%         &        & Figures 1-2\\
% \midrule
% .. \\
% \bottomrule
% \end{tabular}
% \end{center}
% }

\begin{center}
\begin{tabular}{rll}
\toprule
Artifact ID  &  Contributions &  Related \\
             &  Supported     &  Paper Elements \\
\midrule
$A_1$ & $C_1$ & Fig.~1, Fig.~3, Fig.~5--6; \\ & & Table~II; Alg.~2 \\
\midrule
$A_1$ & $C_2$ & Fig.~4; Fig.~7--10 \\
\midrule
$A_1$ & $C_3$ & Fig.~3; Table~III--IV \\
\bottomrule \\
\end{tabular}
\end{center}

%%%%%%%%%%%%%%%%%%%%%%%%%%%%%%%%%%%%%%%%%%%%%%%%%%%%%%%%
\section{Artifact Identification}
%%%%%%%%%%%%%%%%%%%%%%%%%%%%%%%%%%%%%%%%%%%%%%%%%%%%%%%%

\newartifact

\artrel

This artifact is a snapshot of the codebase used in the paper. It lets reviewers run the full workflow (pruned DFS candidate generation, WSC-based selection, and distributed execution) and reproduce the reported results using the provided scripts and logs. Therefore, it directly supports the pruning method ($C_1$), the distributed execution approach ($C_2$), and the empirical evaluation ($C_3$).

% \artexpl{
%     Briefly explain the relationship between the artifact and contributions.
% }

\artexp

Running the artifact should produce the same high-level outcomes reported in the paper: it (i) generates candidate $k$-gene (multi-hit) combinations using P-DFS, (ii) applies the WSC-based selection stage to obtain the final set of combinations, and (iii) reports pruning and runtime summaries (e.g., candidates explored vs.\ pruned and wall-time breakdowns across MPI ranks). These outputs substantiate the main contributions by showing that P-DFS reduces the candidate space relative to a full-enumeration baseline and, as a result, runs faster for $k \ge 3$. Moreover, the advantage should grow with $k$, evidenced by a progressively larger reduction in the number of combinations evaluated.

% \artexpl{
% Provide a higher level description of what outcome to expect from the corresponding experiments. Provide an explanation of how the results substantiate the main contributions.
% }

% \artsampl{
% Algorithm A should be faster than Algorithms C and B in all GPU scenarios.    
% }

\arttime

% \artexpl{
% Estimate the time required to reproduce the artifact, providing separate estimates for the individual steps: Artifact Setup, Artifact Execution, and Artifact Analysis.
% }

% \artsampl{
% The expected computational time of this artifact on GPU X is 20~min.    
% }

The expected reproduction time for this artifact is dominated by execution. Artifact setup (configure/build) takes $<1$~min and artifact analysis (parsing the emitted logs into paper-style summaries) takes $<1$~min. Dataset download and preprocessing typically take 10--15~min once access has been granted (the administrative time to request/obtain access to the TCGA data is not included). Execution time varies with the dataset, the number of hits, and the amount of parallelism (node/rank count). To give context, low hit thresholds can complete quickly (e.g., 2 hits typically finishes in 1--2~min), while higher hits take longer. As a representative reference point from the paper, for the BLCA dataset at 4 hits using P-DFS, execution takes about 133~min on 192 nodes and about 20~min on 3{,}072 nodes. In contrast, without P-DFS (full-enumeration baseline), the BLCA run at 3{,}072 nodes is estimated to take over 52~hours (3{,}120~min).

\artin

\artinpart{Hardware}

We ran the experiments on Fugaku (A64FX CPUs), but the artifact is CPU-only and can run on any Linux workstation or HPC cluster. Parallel runs require an MPI environment (multiple ranks across cores/nodes). No special interconnect is required for correctness; higher-bandwidth networks simply improve scaling at large rank counts. Memory and core requirements depend on the cohort size and the number of genes retained after preprocessing.

\artinpart{Software}

The artifact is the P-DFS-Multihit-WSC repository from https://github.com/RitvikPrabhu/P-DFS-Multihit-WSC. Building/running it requires CMake, a C++17 compiler, an MPI runtime, and Python 3.10+.

\artinpart{Datasets / Inputs}

The artifact uses TCGA somatic mutation files in Mutation Annotation Format (MAF) obtained via the NCI Genomic Data Commons (GDC). Access to TCGA controlled data requires an approved data access request/proposal. This requirement applies to obtaining the MAFs (administrative approval time is external to the artifact). Once access is granted, the MAFs can be downloaded from the GDC Data Portal. The pipeline then preprocesses the MAFs into (i) a combined, bitwise-packed mutation matrix (genes $\times$ samples, tumor and matched-normal) and (ii) a gene map used to decode outputs. The artifact is cohort-agnostic: it can be run on any TCGA cancer type/cohort provided as input in this format.

% \artinpart{Installation and Deployment}

% Clone the repository and build the MPI executable with CMake (optionally Ninja). The build uses compile-time options to set the hit threshold (number of genes per combination) and to enable/disable pruning (P-DFS vs.\ exhaustive baseline). Run the executable with MPI, providing the packed dataset file and output paths for results/metrics. Analysis scripts aggregate the emitted metrics to reproduce paper-style summaries (pruning effectiveness and scaling trends). Exact build/run commands and example workflows are provided in the README of the P-DFS-Multihit-WSC repository.

\artinpart{Installation and Deployment}

To install and build the artifact, clone the repository and compile the MPI executable with CMake (we recommend the Ninja backend). There are two build-time settings you choose up front:

\begin{itemize}\itemsep0.25em
  \item \textbf{Hit threshold (k):} set NUMHITS to the number of genes per combination you want to search for (e.g., NUMHITS=2 for 2-hit, NUMHITS=4 for 4-hit).
  \item \textbf{Mode (pruned vs.\ baseline):} set BOUND=ON to enable pruning (P-DFS) or BOUND=OFF to disable pruning (exhaustive enumeration baseline).
\end{itemize}

A typical configure/build sequence (pruned P-DFS) is:
\begin{quote}\footnotesize
\begin{verbatim}
>> git clone https://github.com/RitvikPrabhu/ \
  P-DFS-Multihit-WSC.git
>> cd P-DFS-Multihit-WSC
>> mkdir -p build && cd build
>> cmake -G Ninja .. -DNUMHITS=<k> -DBOUND=ON
>> ninja
\end{verbatim}
\end{quote}

To build the exhaustive baseline instead, re-run CMake with BOUND=OFF and rebuild:
\begin{quote}\footnotesize
\begin{verbatim}
>> cmake -G Ninja .. -DNUMHITS=<k> -DBOUND=OFF
>> ninja
\end{verbatim}
\end{quote}

This produces the MPI executable used in the experiments. Exact option names and example configurations are provided in the repository README.

% \artinpart{Hardware}

% \artexpl{
% Specify the hardware requirements and dependencies (e.g., a specific interconnect or GPU type is required).
% }

% \artinpart{Software}

% \artexpl{
% Introduce all required software packages, including the computational artifact. For each software package, specify the version and provide the URL.
% }

% \artinpart{Datasets / Inputs}

% \artexpl{
% Describe the datasets required by the artifact. Indicate whether the datasets can be generated, including instructions, or if they are available for download, providing the corresponding URL.
% }

% \artinpart{Installation and Deployment}

% \artexpl{
% Detail the requirements for compiling, deploying, and executing the experiments, including necessary compilers and their versions.
% }

\artcomp

 Assuming the cohort MAF files have already been downloaded from GDC, artifact execution consists of three dependent tasks:
\emph{(1) MAF preprocessing} $\rightarrow$ \emph{(2) merge/pack into solver input} $\rightarrow$ \emph{(3) MPI run}.

\begin{itemize}\itemsep0.25em
  \item \textbf{Task 1: MAF preprocessing (downloaded MAFs $\rightarrow$ intermediates).}
  Run the preprocessing script on the GDC download directory. It scans the MAFs, filters to functionally relevant mutation types, and separates tumor vs.\ matched-normal calls. It produces two intermediate files (written to the current working directory): a tumor mutation matrix and a normal mutation list.
\begin{quote}\footnotesize
\begin{verbatim}
>> python3 utils/preprocessing_maf.py \
    <GDC_COHORT_DIR>
\end{verbatim}
\end{quote}

  \item \textbf{Task 2: Create the combined solver input (intermediates $\rightarrow$ packed input + gene map).}
  Convert the tumor/normal intermediates into the final solver input: a bitwise-packed combined mutation matrix (genes $\times$ samples) and a gene-map file used to decode output indices. This step also sorts genes from most sparse to least sparse.
\begin{quote}\footnotesize
\begin{verbatim}
>> python3 utils/process_gene_data.py \
    <Tumor_matrix_*.txt> \
    <Normal_list_*.txt> \
    <packed_input.bin>
\end{verbatim}
\end{quote}

  \item \textbf{Task 3: MPI execution (packed input $\rightarrow$ metrics + result tuples).}
  After building the executable as described in the Installation and Deployment section, launch the solver with MPI. The run consumes the packed input from Task 2 and creates (i) a metrics file (timing/pruning summaries) and (ii) an output file containing one $k$-hit combination per line as a tuple of gene indices.
\begin{quote}\footnotesize
\begin{verbatim}
>> mpirun -np <ranks> ./run \
    <packed_input.bin> <metrics.txt> \
    <output.txt>
\end{verbatim}
\end{quote}
\end{itemize}

\noindent\textbf{Experimental parameters.} The reproduction-critical parameters are: (i) the cohort directory passed to Task 1, (ii) the hit threshold $k$ (set at compile time via NUMHITS), (iii) pruning mode (set at compile time via BOUND: ON for pruned P-DFS, OFF for exhaustive baseline), and (iv) the parallelism level (MPI rank count in the run command). We report wall time and pruning metrics from the emitted logs. Repetitions are optional. If you want to reduce run to run timing variability you can repeat a configuration three times and report the median.

% \artcomp

% \artexpl{
% Provide an abstract description of the experiment workflow of the artifact. It is important to identify the main tasks (processes) and how they depend on each other. 

% A workflow may consist of three tasks: $T_1, T_2$, and $T_3$. The task $T_1$ may generate a specific dataset. This dataset is then used as input by a computational task $T_2$, and the output of $T_2$ is processed by another task $T_3$, which produces the final results (e.g., plots, tables, etc.). State the individual tasks $T_i$ and provide their dependencies, e.g., $T_1 \rightarrow T_2 \rightarrow T_3$.

% Provide details on the experimental parameters. How and why were parameters set to a specific value (if relevant for the reproduction of an artifact), e.g., size of dataset, number of data points, input sizes, etc. Additionally, include details on statistical parameters, like the number of repetitions.
% }
\artout
Each run writes the two output files that you pass as arguments to the MPI executable. The combinations file stores the selected combinations. It contains one $k$ hit combination per line. The solution itself is the parenthesized tuple of gene indices $(i_1,i_2,\ldots,i_k)$. The tuple length equals $k$. Lines may also include additional per combination fields that indicate the pruning statistics. The gene indices refer to the gene ordering produced during preprocessing rather than gene names. A gene map file is produced during preprocessing and it defines the mapping from index to gene name.

\smallskip
\noindent\textbf{Convert indices to gene names.} To translate tuples into gene names, use the conversion script with the solver output file and the gene map file produced during preprocessing.
\begin{quote}\footnotesize
\begin{verbatim}
>> python3 utils/convertIndexToGeneName.py \
   <output.txt> <gene_map>
\end{verbatim}
\end{quote}

\smallskip
\noindent\textbf{Verify tumor coverage.} To check coverage, use the verification script with the tumor and normal intermediates from preprocessing and the solver output file. The script reports whether the result file achieves the expected tumor coverage and it can list uncovered samples.
\begin{quote}\footnotesize
\begin{verbatim}
>> python3 utils/verifyAccuracy.py \
   <tumor_matrix.txt> <normal_list.txt> \
   <output.txt>
\end{verbatim}
\end{quote}

\smallskip

 Use the output files to report the final selected combinations for each configuration. Use the timing information printed during execution or recorded in the logs to summarize wall time across node and rank counts and across hit thresholds.

 % \clearpage

% \clearpage

% \input{ipdps26_ad_ae.tex}

\end{document}